\documentclass[prb,twocolumn,showpacs,preprintnumbers,amsmath,aps]{revtex4}
\usepackage{graphicx,hyperref}
\usepackage{xcolor}
\usepackage{bm}
\usepackage{subfigure}

\def\nbC{{\mathchoice {\setbox0=\hbox{$\displaystyle\rm C$}%
\hbox{\hbox to0pt{\kern0.4\wd0\vrule height0.9\ht0\hss}\box0}}
{\setbox0=\hbox{$\textstyle\rm C$}\hbox{\hbox
to0pt{\kern0.4\wd0\vrule height0.9\ht0\hss}\box0}}
{\setbox0=\hbox{$\scriptstyle\rm C$}\hbox{\hbox
to0pt{\kern0.4\wd0\vrule height0.9\ht0\hss}\box0}}
{\setbox0=\hbox{$\scriptscriptstyle\rm C$}\hbox{\hbox
to0pt{\kern0.4\wd0\vrule height0.9\ht0\hss}\box0}}}}
\def\nbQ{{\mathchoice {\setbox0=\hbox{$\displaystyle\rm
Q$}\hbox{\raise
0.15\ht0\hbox to0pt{\kern0.4\wd0\vrule height0.8\ht0\hss}\box0}}
{\setbox0=\hbox{$\textstyle\rm Q$}\hbox{\raise
0.15\ht0\hbox to0pt{\kern0.4\wd0\vrule height0.8\ht0\hss}\box0}}
{\setbox0=\hbox{$\scriptstyle\rm Q$}\hbox{\raise
0.15\ht0\hbox to0pt{\kern0.4\wd0\vrule height0.7\ht0\hss}\box0}}
{\setbox0=\hbox{$\scriptscriptstyle\rm Q$}\hbox{\raise
0.15\ht0\hbox to0pt{\kern0.4\wd0\vrule height0.7\ht0\hss}\box0}}}}
\def\nbT{{\mathchoice {\setbox0=\hbox{$\displaystyle\rm
T$}\hbox{\hbox to0pt{\kern0.3\wd0\vrule height0.9\ht0\hss}\box0}}
{\setbox0=\hbox{$\textstyle\rm T$}\hbox{\hbox
to0pt{\kern0.3\wd0\vrule height0.9\ht0\hss}\box0}}
{\setbox0=\hbox{$\scriptstyle\rm T$}\hbox{\hbox
to0pt{\kern0.3\wd0\vrule height0.9\ht0\hss}\box0}}
{\setbox0=\hbox{$\scriptscriptstyle\rm T$}\hbox{\hbox
to0pt{\kern0.3\wd0\vrule height0.9\ht0\hss}\box0}}}}
\def\nbS{{\mathchoice
{\setbox0=\hbox{$\displaystyle     \rm S$}\hbox{\raise0.5\ht0%
\hbox to0pt{\kern0.35\wd0\vrule height0.45\ht0\hss}\hbox
to0pt{\kern0.55\wd0\vrule height0.5\ht0\hss}\box0}}
{\setbox0=\hbox{$\textstyle        \rm S$}\hbox{\raise0.5\ht0%
\hbox to0pt{\kern0.35\wd0\vrule height0.45\ht0\hss}\hbox
to0pt{\kern0.55\wd0\vrule height0.5\ht0\hss}\box0}}
{\setbox0=\hbox{$\scriptstyle      \rm S$}\hbox{\raise0.5\ht0%
\hboxto0pt{\kern0.35\wd0\vrule height0.45\ht0\hss}\raise0.05\ht0%
\hbox to0pt{\kern0.5\wd0\vrule height0.45\ht0\hss}\box0}}
{\setbox0=\hbox{$\scriptscriptstyle\rm S$}\hbox{\raise0.5\ht0%
\hboxto0pt{\kern0.4\wd0\vrule height0.45\ht0\hss}\raise0.05\ht0%
\hbox to0pt{\kern0.55\wd0\vrule height0.45\ht0\hss}\box0}}}}
\def\nbZ{{\mathchoice {\hbox{$\sf\textstyle Z\kern-0.4em Z$}}
{\hbox{$\sf\textstyle Z\kern-0.4em Z$}}
{\hbox{$\sf\scriptstyle Z\kern-0.3em Z$}}
{\hbox{$\sf\scriptscriptstyle Z\kern-0.2em Z$}}}}

\begin{document}

\title{Approach to the lower critical dimension of the $\varphi^4$ theory in the derivative expansion of the Functional Renormalization Group}

\author{Lucija Nora Farka\v{s}} \email{lnf@phy.hr}
\affiliation{Department  of  Physics,  University  of  Zagreb,  Bijeni\v{c}ka  c. 32,  10000  Zagreb,  Croatia}
\affiliation{LPTMC, CNRS-UMR 7600, Sorbonne Universit\'e, 4 Place Jussieu, 75252 Paris cedex 05, France}

\author{Gilles Tarjus} \email{tarjus@lptmc.jussieu.fr}
\affiliation{LPTMC, CNRS-UMR 7600, Sorbonne Universit\'e, 4 Place Jussieu, 75252 Paris cedex 05, France}

\author{Ivan Balog} \email{balog@ifs.hr}
\affiliation{Institute of Physics, P.O.Box 304, Bijeni\v{c}ka cesta 46, HR-10001 Zagreb, Croatia}

\date{\today}

\begin{abstract}
We revisit the approach to the lower critical dimension $d_{\rm lc}$ in the Ising-like $\varphi^4$ theory within the functional renormalization group by studying 
the lowest approximation levels in the derivative expansion of the effective average action. Our goal is to assess how the latter, which provides a generic approximation 
scheme valid across dimensions and found to be accurate in $d\geq 2$, is able to capture the long-distance physics associated with the expected proliferation 
of localized excitations near $d_{\rm lc}$. We show that the convergence of the fixed-point effective potential is nonuniform when $d\to d_{\rm lc}$ with the 
emergence of a boundary layer around the minimum of the potential. This allows us to make analytical predictions for the value of the lower critical dimension 
$d_{\rm lc}$ and for the behavior of the critical temperature as $d\to  d_{\rm lc}$, which are both found in fair agreement with the known results. This confirms the 
versatility of the theoretical approach. 
\end{abstract}


\maketitle

\section{Introduction}
\label{sec:introduction}

Collective behavior characterized by an emergent scale invariance is encountered in a wide variety of physical situations where many degrees of freedom are correlated 
over long distances. Since its introduction, the Renormalization Group has been the theoretical tool of choice for understanding and describing this 
phenomenon.\cite{wilson-kogut} It provides a powerful conceptual framework but, exact results being scarce, the search for generic and efficient approximation schemes 
has been very active from the very beginning.\cite{wilson-wavelet,wilson-fisher72,ma72,migdal-kadanoff} One relatively recent line of 
research starts from an exact formulation of the Renormalization Group, in the form of a functional Renormalization Group (FRG) for scale-dependent generating 
functionals of correlation functions,\cite{wegner-houghton,polchinski,wetterich93} and introduces potentially {\it nonperturbative} approximations through ansatzes for the 
scale-dependent generating functional under study. The question we want to address is to what extent such generic approximation schemes 
are able to describe specific problems in which the long-distance behavior involves strongly nonuniform configurations with localized excitations.

An example of such an approximation scheme within the FRG is the so-called derivative expansion of the effective average action (coarse-grained Gibbs free 
energy in the language of magnetic systems), which amounts to truncating the functional form of the latter in powers of the external momenta or equivalently in 
gradients of the fields.\cite{morris94} The versatility and the effectiveness of the approach have been discussed in several reviews: see Refs.~[\onlinecite{berges02,dupuis21}]. 
One key advantage of such an approach is that space dimension $d$ (as well as number of components of the fields, etc.) can be continuously varied at will, allowing 
one to describe critical behavior from the upper dimension $d_{\rm uc}$ where spatial fluctuations of the local order parameter are easily tamed and classical 
(mean-field) exponents are observed down to the lower critical dimension $d_{\rm lc}$ below which fluctuations become so strong that no phase transition is 
possible. 

The derivative-expansion approximation focuses on the long-distance properties and, in terms of coarse-grained configurations of the system, works about 
uniform configurations. One may therefore wonder if such a scheme is able to capture the physics associated with nonuniform configurations 
containing, {\it e.g.}, domain walls, spin waves, or localized defects. The answer appears to be positive in the case of configurations involving extended 
defects, {\it i.e.}, defects whose energy scales with the system size but in a subextensive way. For instance, the effect 
of spin waves or domain walls which are associated with the return to convexity of the free energy of an O($N$) model in its low-temperature ordered phase 
when spatial fluctuations are taken into account,\cite{ringwald-wetterich,tetradis92,berges02,pelaez-wschebor} or the role of singular avalanche events 
and of scale-free droplet excitations in the critical random-field Ising model\cite{tissier06,tissier12,tarjus20} are all properly accounted for by the truncated 
derivative expansion even at the lowest orders. 

Yet, the jury is still out when the relevant coarse-grained configurations that control the large-scale behavior involve localized excitations such as 
the kinks and anti-kinks found in the instanton analysis of the 1-dimensional Ising model.\cite{rulquin15}  
As the approach to the lower critical dimension for systems with a discrete symmetry is expected to be controlled by the proliferation of such 
localized excitations,\cite{bruce81,bruce83,wallace84} describing the long-distance physics in, say, a model in the Ising universality class such as the $\varphi^4$ 
theory in $d=1+\epsilon$ when $\epsilon\to 0$ is thus a more demanding task for the nonperturbative but approximate FRG than describing the O($N>2$) universality 
class near $d=2$.\cite{footnote_2d}

In this paper we investigate how low orders of the derivative expansion in the FRG describe the approach to the lower critical dimension of the $\varphi^4$ 
theory. The lowest order is known as the Local Potential Approximation (LPA)\cite{wegner-houghton} and is clearly unphysical in low dimensions as it predicts 
$d_{\rm lc}=2$. Indeed, field renormalization is not accounted for in the LPA, implying that the anomalous dimension of the field is always $\eta=0$. This 
then misses a crucial ingredient for investigating Ising criticality in dimensions less than $2$. We thus consider the simplest approximation beyond the 
LPA that incorporates this effect and is often referred to as the LPA'.\cite{berges02,dupuis21} Working at this level allows us to provide a detailed analytical 
treatment of the problem. 

We stress that the issue {\it per se} is not to provide another characterization of the $\varphi^4$ theory near $d=1$, as for instance Bruce and 
Wallace\cite{bruce81,bruce83,wallace84} have already developed an efficient approach in terms a specifically tailored droplet theory. The issue is to assess the 
ability of a {\it generic} nonperturbative approximation scheme within the FRG to quantitatively describe the long-distance physics of a model across the whole 
range of space dimensions from $d_{\rm lc}$ to $d_{\rm uc}$  {\it without a priori knowledge} of the relevant real-space coarse-grained configurations. This also 
involves the question of the continuity of the critical behavior in the dimensionality of space, which was first investigated by Ballhausen et al.\cite{ballhausen04} 
At odds with the latter work we show that convergence of the critical behavior of the $\varphi^4$ theory when $d\to d_{\rm lc}$ within the FRG is nonuniform.

The outline of the paper is as follows. In Sec.~\ref{sec:FRG} we summarize the FRG framework and the derivative expansion scheme for the scalar 
$\varphi^4$ theory. We also introduce the LPA' approximation and the approach to the lower critical dimension $d_{\rm lc}$. We then show in 
Sec.~\ref{sec:nonuniform} that the convergence of the fixed-point effective potential to the lower critical dimension is nonuniform in the field and involves 
a boundary (or interior) layer around the minimum of the potential. We detail the singular perturbation treatment that allows us to find the solution at 
leading order over the whole range of field. We next present in Sec.~\ref{sec:results} the results that we obtain for the value the lower critical 
dimension, which we find close to the exact value $d_{\rm lc}=1$, as well as for the critical temperature and for the critical exponents as $d\to d_{\rm lc}$. We 
finally give some concluding remarks and provide additional details on the method and the solution in several appendices.

\section{Functional RG, derivative expansion, and the LPA'}
\label{sec:FRG}

We are interested in the critical behavior of the Ising universality class, which can be represented at a field-theoretical level by a a scalar $\varphi^4$ theory,
\begin{equation}
\label{eq_bare-action}
S[\varphi]=\int_x \big ( \frac{1}{2}(\partial_x \varphi(x))^2 + \frac{r}{2} \varphi(x)^2 + \frac{u}{4!}\varphi(x)^4 \big ),
\end{equation}
where $\int_x\equiv \int d^d x$. To do so, we use the FRG approach which is a modern version of Wilson's RG in which fluctuations are progressively 
incorporated in the calculation of the partition function of the model through the addition to the action of an infrared (IR) regulator,\cite{berges02}
\begin{equation}
\label{eq_regulator}
\Delta S_k[\varphi]=\frac{1}{2} \int_{xy} R_k(x-y)\varphi(x)\varphi(y),
\end{equation}
where $R_k$ is an IR cutoff function that suppresses integration of modes with momenta less than $k$ without altering that of modes with momenta larger 
than $k$. Typical choices of $R_k$ will be discussed below. The modified partition function
\begin{equation}
Z_k[J]=\int \mathcal D\varphi \exp(-S[\varphi]-\Delta S_k[\varphi])
\end{equation}
is the scale-dependent generating functional of correlation functions and via a modified Legendre transform,
\begin{equation}
\label{eq_Legendre}
\Gamma_k[\phi]=-\ln Z_k[J]+\int_x J(x)\phi(x)-\Delta S_k[\phi]),
\end{equation}
one can introduce the effective average action $\Gamma_k[\phi]$, with $\phi(x)=\langle \varphi(x)\rangle=\delta \ln Z_k[J]/\delta J(x)$, which is the 
scale-dependent generating functional of the 1-particle irreducible (1-PI) correlation functions. It obeys an exact functional RG equation that 
describes its evolution with the IR scale $k$,\cite{wetterich93}
\begin{equation}
\label{eq_ERGE}
\partial_t\Gamma_k[\phi]=\frac 12\int_{xy}\partial_t R_k(x-y)\big [(\Gamma_k^{(2)}[\phi]+R_k)^{-1}\big ]_{xy},
\end{equation}
where $\Gamma_k^{(2)}$ is the second functional derivative of $\Gamma_k$ and $t=\ln(k/\Lambda)$ with $\Lambda$ a UV cutoff.

The exact FRG equation in Eq.~(\ref{eq_ERGE}) is a convenient starting point for devising nonperturbative approximation schemes in the form of ansatzes for 
the functional dependence of the effective average action. One such scheme used to capture the long-distance physics is the derivative expansion in which 
the Lagrangian associated with $\Gamma_k$ is expanded in gradients of the fields,
\begin{equation}
\label{eq_DE}
\Gamma_k[\phi]=\int_x \big [ U_k(\phi(x))+\frac{1}{2}Z_k(\phi(x))(\partial_x \phi(x))^2 + {\rm O}(\partial\partial\partial\partial) \big ].
\end{equation}
When inserted in Eq.~(\ref{eq_ERGE}) the above ansatz provides a hierarchy of coupled FRG equations for the functions $U_k(\phi)$, $Z_k(\phi)$, etc., where 
the field configurations involved are now uniform, {\it i.e.}, $\phi(x)=\phi$.

Scale invariance associated with criticality is described by a fixed point in the FRG equations once the latter have been cast in a dimensionless form via the use 
of scaling dimensions. One defines dimensionless quantities $\varphi$, $u_k$, $z_k$, etc., through
\begin{equation}
\phi=k^{D_\phi} \varphi,\;\; U_k(\phi)=k^du_k(\varphi),\;\; Z_k(\phi)=Z_k z_k(\varphi),
\end{equation}
etc., where the dimension of the field is related to the anomalous dimension $\eta$ by $D_\phi=(d-2+\eta)/2$ and where the field renormalization constant 
$Z_k$ goes as $k^{-\eta}$ in the vicinity of the fixed point. (Note that we have used the same notation $\varphi$ for the bare variable in Eq.~(\ref{eq_bare-action}) 
and the dimensionless average field, as the former will no longer appear in what follows.)

The hierarchy of FRG equations when expressed in terms of dimensionless quantities takes the form
\begin{equation}
\begin{aligned}
\label{eq_ERGEdimensionless}
&\partial_t u_k(\varphi)=-d u_k(\varphi) +\frac{(d-2+\eta_k)}{2}\varphi u_k'(\varphi) + \beta_{u}(\varphi;\eta_k) \\&
\partial_t z_k(\varphi) = \eta_k z_k(\varphi) +\frac{(d-2+\eta_k)}{2}\varphi z_k'(\varphi) + \beta_{z}(\varphi;\eta_k),
\end{aligned}
\end{equation}
etc., where a prime indicates a derivative with respect to the argument of the function; $\beta_u$, $\beta_z$, etc. are functionals of $u_k''$, $z_k$, etc., 
and are given in Appendix~\ref{appx_flows}. The fixed points of the flow equations are reached when $t\to -\infty$ ({\it i.e.}, $k\to 0$) and the left-hand sides go to 
zero.

As already stressed, a proper description of the approach to the lower critical dimension should incorporate field renormalization and a nonzero anomalous 
dimension $\eta$. The lowest order of the derivative expansion that achieves this is the so-called LPA' in which one retains on top 
of the renormalized potential $U_k(\phi)$ a field independent but scale dependent $Z_k$. In explicit form, the dimensionless equation for the fixed-point 
potential is now
\begin{equation}
\begin{aligned}
\label{eq_FPpotential}
&0=-d\, u(\varphi) +\frac{d-2+\eta}2 \varphi u'(\varphi) +2 v_d \ell_0^{(d)}(u''(\varphi);\eta),
\end{aligned}
\end{equation}
where $v_d^{-1}=2^{d+1} \pi^{d/2}\Gamma(d/2)$ and $\ell_0^{(d)}$ is a (strictly positive) dimensionless threshold function which enforces the decoupling 
of the low-momentum and high-momentum modes; it is defined in terms of the dimensionless IR cutoff function $r(y=q^2/k^2)=R_k(q^2)/(Z_k q^2)$ by
\begin{equation}
\begin{aligned}
\label{eq_threshold_ell0}
\ell_0^{(d)}(w;\eta)= -\frac 12\int_0^{\infty}dy y^{\frac d2}\frac{\eta r(y)+2yr'(y)}{(y[1+r(y)]+w)}
\end{aligned}
\end{equation}
and is described in more detail in Appendix~\ref{app:threshold}. We have dropped the subscript $k\to 0$ for quantities at the fixed point in the above 
equation to simplify the notation.

Deriving Eq.~(\ref{eq_FPpotential}) gives an equation for $u'(\varphi)$ from which one extracts the equation for its minima $\pm \varphi_{\rm m}$ [through 
$u'(\pm \varphi_{\rm m})=0$],
\begin{equation}
\label{fl_eq_phim}
0=\frac{(d-2+\eta)}{2}\varphi_{{\rm m}}+2v_d\frac{u'''(\varphi_{\rm m})}{u''(\varphi_{{\rm m}})}\partial_w\ell_0^{(d)}(w;\eta)\vert_{w=u''(\varphi_{\rm m})},
\end{equation}
and deriving one more time gives an equation for the ``squared mass'' $u''(\varphi)$. Both equations will be useful below.

In the LPA' the field renormalization constant $Z_k$ is chosen such that at the minimum of the potential $z_k(\pm \varphi{\rm m})=1$.\cite{berges02,ballhausen04} 
From Eq.~(\ref{eq_ERGEdimensionless}) and the explicit form of $\beta_z$ given in Appendix~\ref{appx_flows} one then obtains that
\begin{equation}
\label{eq_anomalous-dim}
\eta= \frac{4v_d}d u'''(\varphi_{\rm m})^2 m_{4,0}^{(d)}(u''(\varphi_{\rm m});\eta),
\end{equation}
where $m_{4,0}^{(d)}$ is another (strictly positive) dimensionless threshold function defined by
\begin{equation}
\begin{aligned}
\label{eq_threshold_m40}
&m_{4,0}^{(d)}(w;\eta)= \frac 12 \int_0^{\infty}dy y^{\frac d2} \frac{1+(yr(y))'}{(y[1+r(y)]+w)^4}
\bigg [2\eta (yr(y))'\\&
+4(y^2r'(y))' -4 \,\frac {y[1+(yr(y))'][\eta r(y)+2yr'(y)]}{y[1+r(y)]+w}  \bigg ]
\end{aligned}
\end{equation}
and discussed in Appendix~\ref{app:threshold}. Once a specific form for the dimensionless IR cutoff function $r(y)$ has been 
chosen, the solution of Eqs.~(\ref{eq_FPpotential}-\ref{eq_threshold_m40}) fully characterizes the LPA' fixed point. In what follows we will use 
two much studied forms of $r(y)$:
\begin{equation}
\begin{aligned}
\label{eq_cutoff_choices}
&r(y)=\alpha \Theta(1-y)(1-y)/y \\&
r(y)=\alpha e^{-y}/y
\end{aligned}
\end{equation}
where $\Theta$ is the Heaviside step function and $\alpha$ is a variational parameter of O(1) that can be determined, {\it e.g.}, by the principle of minimum 
sensitivity.\cite{litim01,dupuis21,balog19} We will refer to these two choices as Theta and Exponential cutoff functions.

\begin{figure}
\begin{center}
\begin{picture}(230,330)
\put(0,170){\includegraphics[width=220pt]{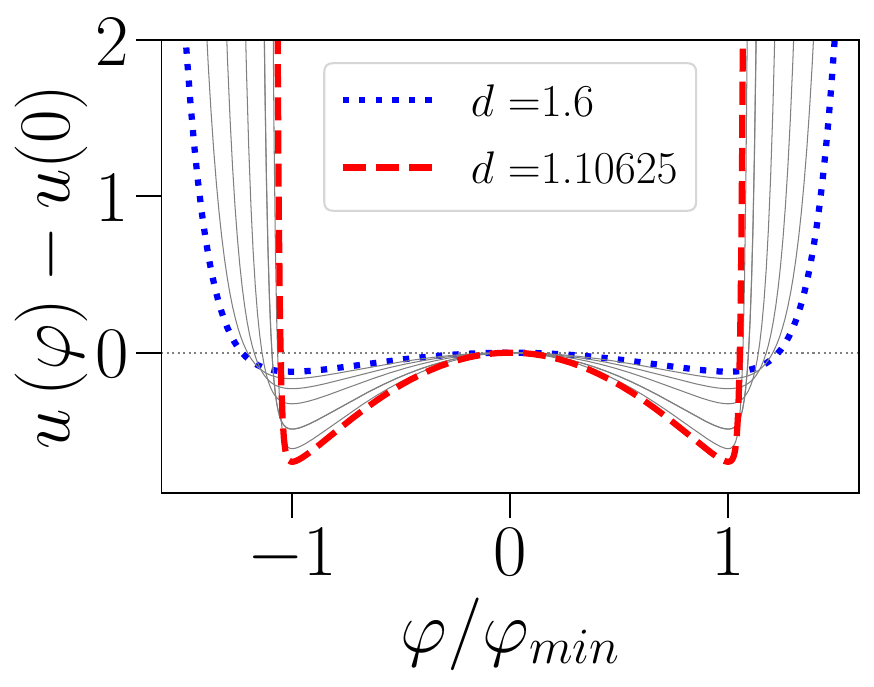}}
\put(52,225){{\huge a)}}
\put(0,0){\includegraphics[width=220pt]{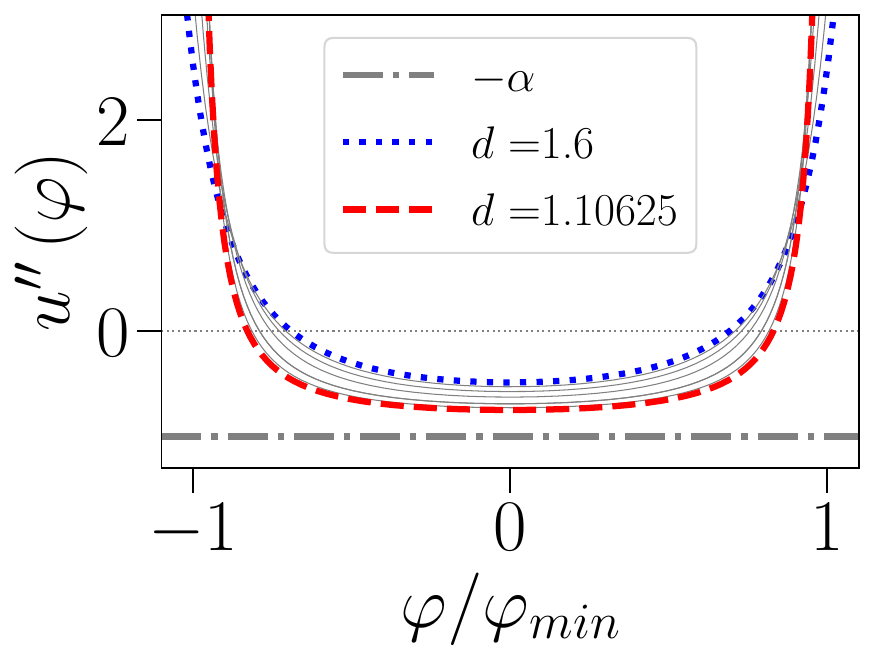}}
\put(52,60){{\huge b)}}
\end{picture}
\end{center}
\caption{Dimensionless effective potential $u(\varphi)$ (a) and its second derivative $u''(\varphi)$ (b) at the LPA' fixed point for several dimensions $d$ between 
$1.6$ and $1.1$. We have used the Theta IR cutoff function with $\alpha=1$ and a numerical resolution of the FRG equations. }
\label{Fig_potential_LPA'}
\end{figure}

\begin{figure}
\includegraphics[width=0.9\linewidth]{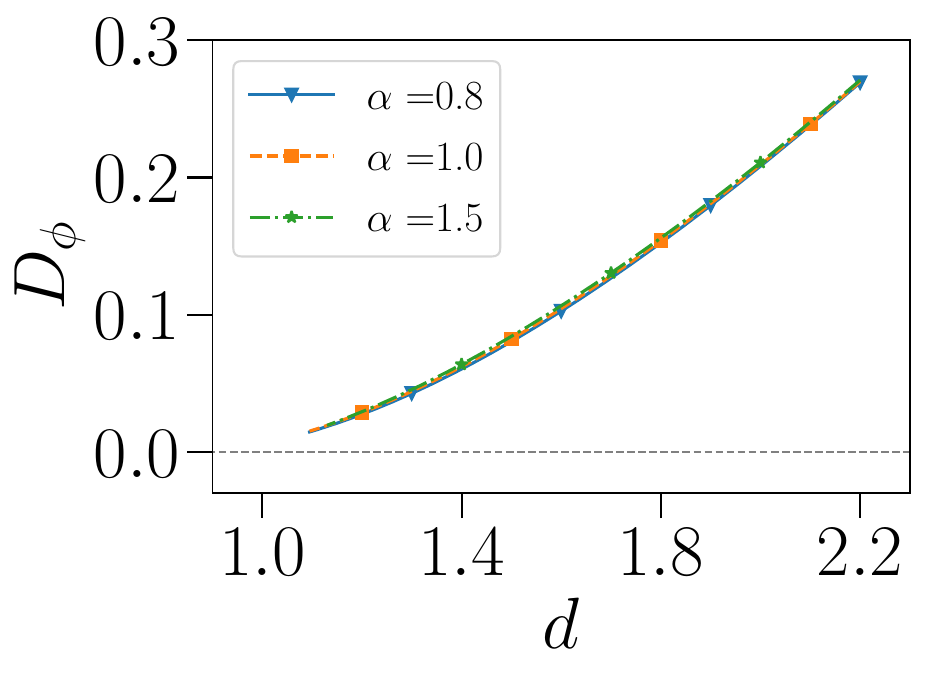}
\caption{Variation with the space dimension $d$ of the field scaling dimension $D_\phi=(d-2+\eta)/2$ at the LPA' fixed point as obtained from a numerical resolution 
and the Exponential IR cutoff function.} 
\label{Fig_field-dimension_LPA'}
\end{figure}

\begin{figure}
\begin{center}
\begin{picture}(230,320)
\put(-5,160){\includegraphics[width=235pt]{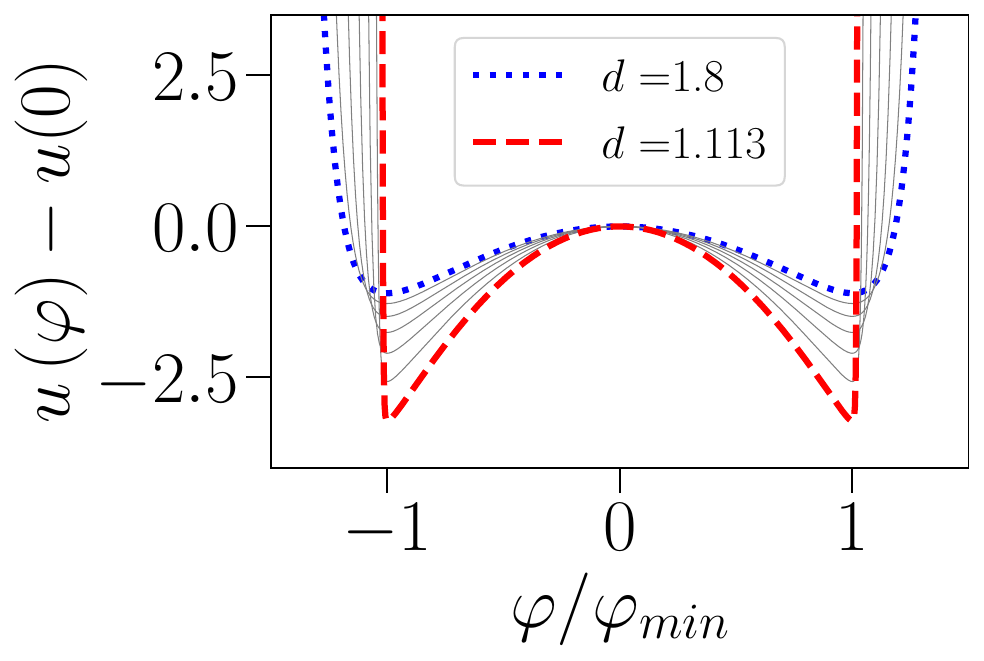}}
\put(65,215){{\huge a)}}
\put(0,0){\includegraphics[width=225pt]{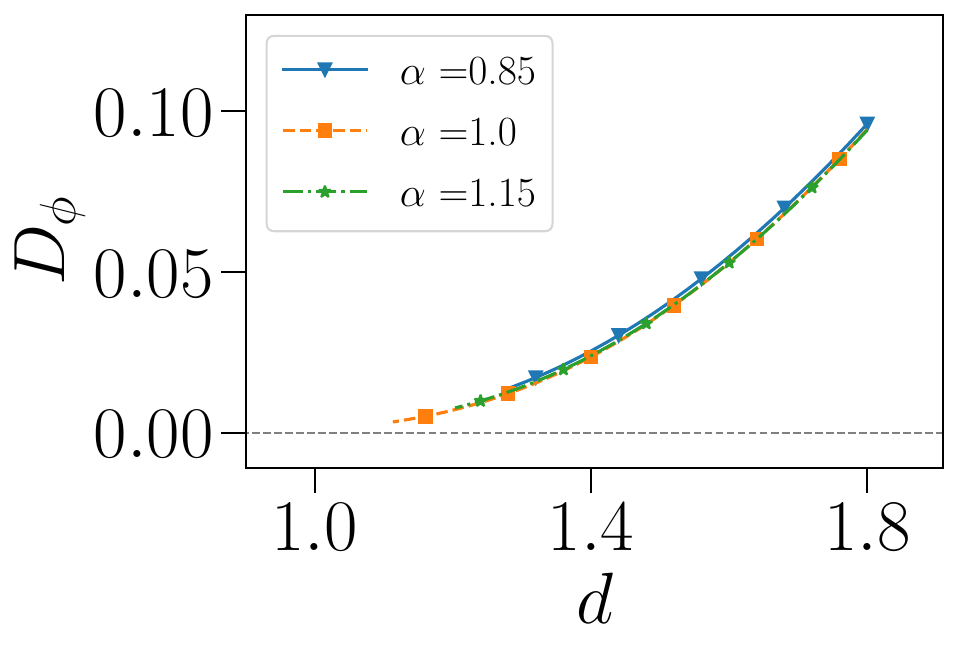}}
\put(65,60){{\huge b)}}
\end{picture}
\end{center}
\caption{Dimensionless effective potential $u(\varphi)$ (a) and field scaling dimension $D_\phi=(d-2+\eta)/2$ (b) at the fixed point obtained of the second order 
of the derivative expansion for several dimensions $d$ between $1.8$ and $1.11$. We have used the Exponential  IR cutoff function with $\alpha=1$ and a 
numerical resolution of the FRG equations.}
\label{Fig_de2}
\end{figure}

We illustrate the results for the fixed point at LPA' and the choice of the Theta cutoff function with $\alpha=1$ (similar results are obtained with other 
choices, see the discussion further below) in Figs.~\ref{Fig_potential_LPA'} and \ref{Fig_field-dimension_LPA'}. Fig~\ref{Fig_potential_LPA'} displays the evolution 
of the dimensionless potential $u(\varphi)$ and the ``square mass'' function $u''(\varphi)$ as the space dimension $d$ decreases, and 
Fig.~\ref{Fig_field-dimension_LPA'} that of the field scaling dimension $D_\phi=(d-2+\eta)/2$. To show that a similar behavior in low dimension is also 
expected for higher orders of the derivative expansion, so that the LPA' level is not atypical, we plot the evolution with $d$ of the dimensionless potential and 
of the field dimension $D_\phi$ at the second order of the derivative expansion for which the field renormalization is now a full function of the field 
[see Eq.~(\ref{eq_ERGEdimensionless})] in Fig.~\ref{Fig_de2}

A defining property of the lower critical dimension is that $(d-2+\eta)|_{d\to d_l}\to 0$. This is equivalent to stating that the scaling dimension $D_\phi$ of the field 
vanishes (see above and Fig.~\ref{Fig_field-dimension_LPA'}). If the field does not rescale, its fluctuations along the RG flow remain of order $1$ in terms of the 
dimensionful field and ordering associated with a nonzero dimensionful average field in zero applied source is impossible. We thus find it convenient to define 
\begin{equation}
\label{eq_tildeepsilon_def}
\tilde\epsilon(d)=\frac{d-2+\eta}{2(2-\eta)},
\end{equation}
which goes to $0$ as $d\to d_{\rm lc}$. We use the notation $\tilde\epsilon$ to avoid confusion with $\epsilon=d-d_{\rm lc}$ (with $d_{\rm lc}=1$ in an exact treatment.)

Another anticipated feature of the approach to the lower critical dimension is the fact that the propagator of the theory approaches a pole. Indeed, the lower 
critical dimension corresponds to the merging of the critical fixed point and the zero-temperature fixed point associated with the symmetry-broken 
ordered phase, and the return to convexity of the effective potential along the FRG flow is controlled in the latter by the presence of a 
pole in the propagator.\cite{tetradis92,berges02,pelaez-wschebor} In the LPA' the dimensionless propagator is given by
\begin{equation}
\label{eq_dimless-propag}
p(y;\varphi)=\frac{1}{y[1+ r(y)]+u''(\varphi)}
\end{equation}
and must of course be positive. With the choice of the Theta cutoff function in Eq.~(\ref{eq_cutoff_choices}) the pole in either in $y=q^2/k^2=0$ and 
$u''(\varphi)=-\alpha$ when $\alpha<1$ or in $y=1$ and $u''(\varphi)=-1$ when $\alpha>1$. With the Exponential cutoff function the pole is either in 
$y=0$ and $u''(\varphi)=-\alpha$ when $\alpha<1$ or in $y=\ln\alpha$ and $u''(\varphi)=-(1+\ln\alpha)$ when $\alpha>1$: see Appendix~\ref{app:threshold}.

In the following we will investigate in more detail the structure of the fixed-point solution at LPA' when $d\to d_{\rm lc}$. As the numerical solution becomes 
harder if not impossible in this limit, progress should be made through an analytical treatment. We stress again that contrary to what 
happens for the O($N>2$) models where there are Goldstone modes associated with the breaking of a continuous symmetry which quite straightforwardly imply 
that $d_{\rm lc}=2$,\cite{zinn-justin89,berges02,delamotte04,dupuis21} the LPA' or any level of truncation of the derivative expansion within 
the FRG need not predict the exact value of the lower critical dimension, $d_{\rm lc}=1$, for the present model with a discrete $Z_2$ symmetry. The 
approximate $d_{\rm lc}$ must then be computed.

\section{Nonuniform convergence to the lower critical dimension}
\label{sec:nonuniform}

\subsection{nonuniform convergence and boundary layer}
\label{sub:BL}

Consider the LPA' fixed-point equation for the second derivative of the potential,
\begin{equation}
\begin{aligned}
\label{eq_derivative-FPpotential}
&0=- u''(\varphi) + \tilde\epsilon(d)\varphi u'''(\varphi) +\frac{2 v_d}{2-\eta(d)} \partial^2_\varphi\ell_0^{(d)}(u''(\varphi);\eta(d)),
\end{aligned}
\end{equation}
where $\eta= 2-d +d\tilde\epsilon+{\rm O}(\tilde\epsilon^2)$. As the dependence of $\tilde\epsilon$ on $d$ is expected to be monotonic, one can study the above 
equation at fixed $\tilde\epsilon$ instead of fixed $d$, and when $\tilde\epsilon\to 0$,
\begin{equation}
\begin{aligned}
\label{eq_derivative-FPpotential_tilde}
0=- u''(\varphi) + \tilde\epsilon\varphi u'''(\varphi) +\frac{2 v_d}{d} \partial^2_\varphi\ell_0^{(d)}(u''(\varphi);2-d),
\end{aligned}
\end{equation}
where $d\equiv d(\tilde\epsilon)\to d_{\rm lc}$ and $d_{\rm lc}$ {\it a priori} unknown.

If $u''(\varphi)$, its first derivatives, and $\varphi$ are of O(1) the second term in Eq.~(\ref{eq_derivative-FPpotential_tilde}) can be set to zero as a 
leading approximation in which $\tilde\epsilon=0$ altogether. This describes a uniform convergence toward the lower critical dimension, as  was assumed in Ref.~[\onlinecite{ballhausen04}]. Our claim, which we substantiate below, is that the second term leads to a singular-perturbation problem and that the 
limit $\tilde\epsilon\to 0$ is nonuniform in the field.

From the shape of the fixed-point potential in Fig.~\ref{Fig_potential_LPA'} one can see that three domains of field values can be distinguished: the large-field 
region, $\vert\varphi\vert\to +\infty$, where $u(\varphi)$ and its first derivatives blow up, the region of the minima where the first derivative is zero but higher-order 
derivatives grow large as $d$ decreases, and the inner region of fields of order O(1) in which the potential and its derivatives appear of O(1). The latter region 
should be describable by the $\tilde\epsilon=0$ equation (plus regular perturbation in $\tilde\epsilon$) and the large-field one corresponds to the situation where 
the square mass $u''(\varphi)$ diverges and the nontrivial beta functions go to zero to only leave the scaling part of the equation: here, 
$0=- u''(\varphi) + \tilde\epsilon\varphi u'''(\varphi)$, which leads to
\begin{equation}
u''(\varphi)\sim \vert\varphi\vert^{\frac 1{\tilde\epsilon}}\;\;{\rm when}\;\; \varphi \to\pm\infty.
\end{equation}
The region of the close vicinity of the minima needs more care and entails a boundary-layer treatment. (Note that interior layer would be more adequate in this 
case than boundary layer because the region in which variation is very fast is away from the boundaries, but with this caveat we will nonetheless keep using 
the term boundary layer.) A unique global solution valid for all fields is finally obtained by ``matching'' the partial solutions obtained in each domain in the 
intermediate regions of field over which they overlap. This matching procedure is a key element of the singular perturbation treatment.\cite{singular-perturbation}

The potential $u(\varphi)$ being $Z_2$ symmetric, we choose to restrict our analysis to positive fields, $\varphi\geq 0$.

\subsection{The solution of the $\tilde{\epsilon}=0$ equation}
\label{sub:outer0}

Consider first the LPA' equation for $\tilde{\epsilon}=0$. Introducing for simplicity the notation $w(\varphi):=u''(\varphi)$, one has
\begin{equation}
\begin{aligned}
\label{eq_epsilon-zero}
w(\varphi) =\frac{2 v_d}d \partial^2_\varphi\ell_0^{(d)}(w(\varphi);2-d),
\end{aligned}
\end{equation}
where $d=d_{\rm lc}$, which we assume in the following to be strictly less than $2$, and the initial conditions are $w(0)=w_0$ and all the 
odd derivatives of $w$ are zero in $\varphi=0$. Let also define $\Phi(\varphi):=\ell_0^{(d)}(w(\varphi);2-d)$. The function $\ell_0^{(d)}(w;2-d)$ 
being monotonically increasing with $w$, one can invert it and define $w=F(\Phi)$ with $F$ such that $F(\ell_0^{(d)}(w;2-d)=w$. Eq.~(\ref{eq_epsilon-zero}) 
can then be rewritten as
\begin{equation}
\begin{aligned}
\label{eq_epsilon-zero2}
\partial^2_\varphi\Phi(\varphi)=\frac d{2 v_d} F(\Phi(\varphi)),
\end{aligned}
\end{equation}
which is the equation of motion of an anharmonic oscillator $\Phi(\varphi)$ with $\varphi$ playing the role of time and $(d/(2v_d))F(\Phi)$ 
being the force. The solution for $\Phi(\varphi)$ is a periodic function starting in $\Phi_0=\ell_0^{(d)}(w_0;2-d)$ with a velocity 
$\partial_\varphi\Phi\vert_0=0$. The half-period $\varphi_*$ corresponds to the first time at which the velocity is again equal to $0$. By using the 
energy balance equation associated with Eq.~(\ref{eq_epsilon-zero2}),
\begin{equation}
\begin{aligned}
\label{eq_epsilon-zero3}
\int_{\Phi_0}^{\Phi(\varphi)}d\Phi' \frac d{2 v_d} F(\Phi') =\frac 12 [\partial_\varphi \Phi(\varphi)]^2,
\end{aligned}
\end{equation}
one derives that $\varphi_*$ is obtained from
\begin{equation}
\begin{aligned}
\label{eq_epsilon-zero_min}
\int_{0}^{\varphi_*}d\varphi' \partial_{\varphi'}\Phi(\varphi')w(\varphi') =0.
\end{aligned}
\end{equation}
Note that $\varphi_*$ and the solution $\Phi(\varphi)$ are parametrized by the initial value $w_0$.

Because of the monotonic relation between $\Phi$ and $w$, the solution for $w(\varphi)$ is also a periodic function of half-period $\varphi_*$ that oscillates 
between a minimum value $w_0$ and a maximum one $w_*=w(\varphi_*)$, the two values being uniquely related. Clearly, this solution cannot be that of the full 
problem (which is not periodic: see Fig.~\ref{Fig_potential_LPA'}(b)) when $\varphi$ is close to and larger than $\varphi_*$. As alluded to above, a 
boundary-layer type of solution must replace the solution of the $\tilde\epsilon=0$ equation. Since $w(\varphi)$ is very large in the close vicinity of the (exact) 
minimum of the potential,  $\varphi_{\rm m}$, a potential matching between the two types of solution must take place for $\varphi<\varphi_{\rm m}\lesssim \varphi_*$, 
which  requires that $w(\varphi_*)$ very large. In this limit, it can be shown from the properties of the $\tilde\epsilon=0$ solution (see Appendix~\ref{app:phi_star}) 
that
\begin{equation}
\label{eq_phi_star}
\varphi_*\sim \sqrt{\ln w_*}.
\end{equation}
The matching requirement and the constraint it puts on the value of $w(\varphi_0)=w_0$ will be considered in more detail below.

\subsection{The inner solution within the layer}
\label{sub:inner-solutionBL}

The $\tilde\epsilon=0$ equation ceases to be the proper description when the second term of Eq.~(\ref{eq_derivative-FPpotential_tilde}) becomes of the same order 
as the other terms and  a new solution must be found. Guided by the numerical solution and by physical intuition, we argue that a new solution takes place within 
a ``boundary layer'' (actually, an ``interior layer'': see the above comment) around the minimum $\varphi_{\rm m}$ of the potential. In this region, $w\gg 1$, and 
Eq.~(\ref{eq_derivative-FPpotential_tilde}) 
becomes
\begin{equation}
\begin{aligned}
\label{eq_w_tilde}
0=- w(\varphi) + \tilde\epsilon\varphi w'(\varphi) +\frac{2 v_d}{d}\alpha A_d \partial^2_\varphi \left[\frac 1{w(\varphi)}\right],
\end{aligned}
\end{equation}
where we have used that for large square mass $w$, Eq.~(\ref{eq_threshold_ell0}) gives
\begin{equation}
\begin{aligned}
\label{eq_ell0-asympt}
\ell_0^{(d)}(w;2-d)\sim \frac{\alpha A_d}{w} + {\rm O}(\frac 1{w^{2}})
\end{aligned}
\end{equation}
with, after some manipulations, $A_d$ obtained as
\begin{equation}
\begin{aligned}
\label{eq_Ad}
A_d= d \int_0^{\infty}dy y^{\frac d2}\left [\frac{r(y)}{\alpha}\right ].
\end{aligned}
\end{equation}
More details are given in Appendix~\ref{app:threshold}. It is then convenient to rescale the field $\varphi$ by a multiplicative factor 
$\sqrt{d/(2\alpha v_dA_d)}$, so that Eq.~(\ref{eq_w_tilde}) reads
\begin{equation}
\begin{aligned}
\label{eq_w_tilde-scaled}
0=- w(\varphi) + \tilde\epsilon\varphi w'(\varphi) +\partial^2_\varphi \left[\frac 1{w(\varphi)}\right],
\end{aligned}
\end{equation}
with no explicit dependence on $d$ (which we recall should be taken as $d_{\rm lc}$ in the limit $\tilde\epsilon\to 0$ that we consider).

In the vicinity of $\varphi_{\rm m}$ we introduce a rescaled variable $x=(\varphi-\varphi_{\rm m})/\delta(\tilde\epsilon)$ with $x={\rm O}(1)$ as 
$\tilde\epsilon\to 0$. By balancing the first two terms of Eq.~(\ref{eq_w_tilde-scaled}) one obtains that $\delta(\tilde\epsilon)=\tilde\epsilon\varphi_{\rm m}$, 
{\it i.e.},
\begin{equation}
\label{eq_scaled-variable_BL}
x=\frac{\varphi-\varphi_{\rm m}}{\tilde\epsilon\varphi_{\rm m}},
\end{equation}
where we assume for now, and check later on, that $\tilde\epsilon\varphi_{\rm m}\to 0$ when $\tilde\epsilon\to 0$. By also requiring that the third term 
is of the same order of magnitude as the first two, one is further led to introduce a function $g(x)$ which is defined by
\begin{equation}
\label{eq_def-g_BL}
w(\varphi)=\frac{g(x)}{\tilde\epsilon\varphi_{\rm m}}
\end{equation}
and which is of O(1) when $x$ is of O(1).
The LPA' equation in the boundary layer can then be expressed as
\begin{equation}
\label{eq_BLg}
-g(x)+g'(x)+\partial^2_x\left [\frac 1{g(x)} \right ]=0.
\end{equation}
This is complemented by the equation for the minimum, Eq.~(\ref{fl_eq_phim}), which leads to
\begin{equation}
\label{eq_BLg_min}
g'(0)=g(0)^3.
\end{equation}

The boundary layer equation can be solved in an implicit form by first introducing the auxiliary function
\begin{equation}
\label{eq_Xg}
X(x)=g(x) + \partial_x\left [\frac 1{g(x)}\right ]
\end{equation}
which satisfies $X(0)=0$ because of Eq.~(\ref{eq_BLg_min}). Then, one has that
\begin{equation}
g(x)=X'(x),
\end{equation}
and the solution for $g(x)$ as a function of $X(x)$ is given by
\begin{equation}
\begin{aligned}
\label{eq_sol_BLg}
g(x)=g(0)\frac{e^{-\frac{X(x)^2}{2}}}{\left [1-\sqrt{\frac{\pi}{2}}g(0) +g(0) \int_{X(x)}^{+\infty} dt \,e^{-\frac{t^2}{2}}\right ]}\,.
\end{aligned}
\end{equation}

The interest of the above expression is that it allows us to study the limit $x\to +\infty$ and require matching with the outer solution at large field, 
$w(\varphi\to\infty)\sim \varphi^{1/\tilde\epsilon}\to +\infty$, already derived. (This is the standard method of matched asymptotic expansions used in 
singular perturbation problems.) Choosing the matching region such that $1/\tilde\epsilon\gg x\gg 1$, 
the latter then imposes that $g(x)$ diverges as $\exp(x)$ at large positive $x$. From Eqs.~(\ref{eq_sol_BLg}) and (\ref{eq_Xg}) it is straightforward 
to see that one must have
\begin{equation}
\label{eq_g0min}
1-\sqrt{\frac{\pi}{2}}g(0)=0,
\end{equation}
which fixes $g(0)$ and via Eq.~(\ref{eq_BLg_min}) $g'(0)$.

\begin{figure}
\includegraphics[width=0.8\linewidth]{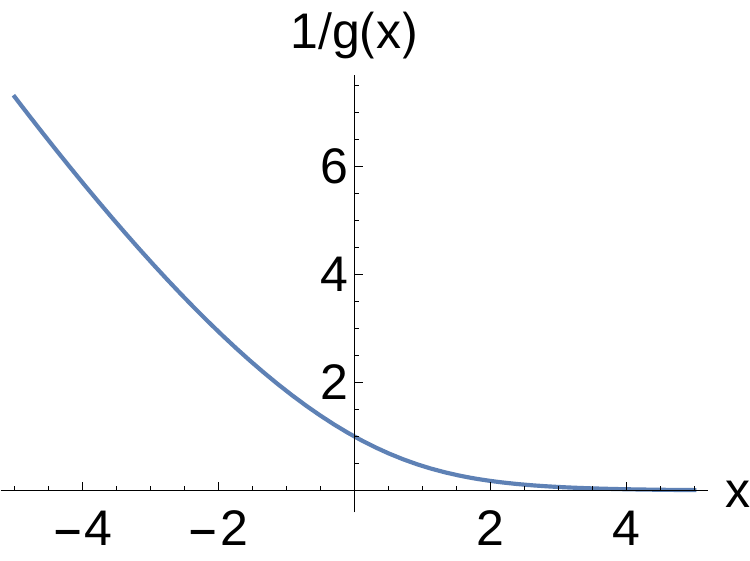}
\caption{Inverse of the inner solution $g(x)$ giving the square-mass function within the layer around the minimum of the potential. The latter corresponds to $x=0$.}
\label{Fig_inner_BL}
\end{figure}

We plot in Fig.~\ref{Fig_inner_BL} the inverse of $g(x)$, as obtained from the solution of the above equations. It is a monotonically decreasing function. 
When $x$ is negative and $\vert x\vert$ very large, it behaves as
one finds that 
\begin{equation}
\begin{aligned}
\label{eq_sol_BLg_minusinfty}
\frac 1{g(x)}\sim \sqrt2 \vert x\vert \sqrt{\ln \vert x\vert}\Big [1+{\rm O}\left (\frac{\ln(\ln \vert x\vert)}{\ln \vert x\vert}\right )\Big ],
\end{aligned}
\end{equation}
an expression which will be useful when considering matching with the outer solution obtained from the equation with $\tilde\epsilon=0$ (see 
Sec.~\ref{sub:outer0}).

\subsection{Matching the inner solution with that of the $\tilde\epsilon=0$ equation}
\label{sub:matching}

At least at the present LPA' level of approximation the solution $g(x)$ within the layer around $\varphi_{\rm m}$ is fully determined by 
matching with the outer solution obtained at large field. However, the relation between $\varphi_{\rm m}$ and $\tilde\epsilon$ is yet to be determined. 
It is also crucial to check that matching with the outer solution corresponding to fields less than $\varphi_{\rm m}$ and to the $\tilde\epsilon=0$ 
equation can be enforced, so that a solution can be constructed for all field values.

In Sec.~\ref{sub:outer0} we have argued that matching should take place for fields $\varphi<\varphi_{\rm m}\lesssim \varphi_*$, where furthermore 
the outer solution associated with the $\tilde\epsilon=0$ equation should be such that $1\ll w(\varphi)\ll  w(\varphi_{\rm m}), w_*$. We 
can choose the matching region where the two solutions overlap such that 
\begin{equation}
\varphi_{\rm m}-\varphi={\rm O}((\tilde\epsilon\varphi_{\rm m})^{a}) \;\; {\rm with}\; 0<a<1,
\end{equation}
and, as a result, $x\sim -(\tilde\epsilon\varphi_{\rm m})^{1-a}$ is negative and very large. The asymptotic limit of the inner solution is then given by 
Eq.~(\ref{eq_sol_BLg_minusinfty}), which implies that
\begin{equation}
w_*\gg w(\varphi)={\rm O}\big (\tilde\epsilon\varphi_{\rm m})^{-a} \vert \ln(\tilde\epsilon\varphi_{\rm m})\vert^{-\frac 12}\big )\to +\infty.
\end{equation}

From Eq.~(\ref{eq_epsilon-zero_min}) one can obtain the relation between $w_*$ and $w_0$ as
\begin{equation}
\begin{aligned}
\label{eq_w*}
0&=\int_{w_0}^{w_*}dw' w' \partial_{w'}\Phi(w')\\&
=w_*\Phi(w_*)-w_0\Phi(w_0)-\int_{w_0}^{w_*}dw' \Phi(w'),
\end{aligned}
\end{equation}
where, we recall, $\Phi(w)=\ell_0^{(d)}(w;2-d)$ with the latter given in Eq.~(\ref{eq_threshold_ell0}). To evaluate the quantities in the above 
equation we split the integral by introducing an intermediate value $w_c$, which we choose  positive and of O(1). Taking into account that $w_*$ diverges 
and using the property that the function $\Phi(w)$ is monotonically decreasing and asymptotically goes to zero as $\alpha A_d/w+ {\rm O}(1/w^2)$, 
we transform Eq.~(\ref{eq_w*}) into 
\begin{equation}
\begin{aligned}
\label{eq_w*2}
\alpha A_d\ln w_*=-w_0\Phi(w_0)+\int_{w_0}^{w_c}dw' \Phi(w'),
\end{aligned}
\end{equation}
up to O(1) terms, and, since $w_0$ and $w_c$ are of O(1), this implies that $\Phi(w_0)$ diverges. This can only occur if $w_0$ approaches the pole of 
the propagator, which we call $w_P$ and can be either $-1$, $-\alpha$, or $-(1+\ln\alpha)$ depending on the IR cutoff function and on $\alpha$ [see 
Eq.~\ref{eq_dimless-propag}) and below]. Notwithstanding the precise asymptotic behavior of $\Phi(w)\equiv\ell_0^{(d)}(w;2-d)$ when 
the pole is approached (this depends on the IR cutoff function, see Appendix~\ref{app:threshold}), the second term of the right-hand side is 
subdominant compared to the first one and one has
\begin{equation}
\begin{aligned}
\label{eq_w*3}
\Phi(w_0)\sim\frac{\alpha A_d}{\vert w_P\vert}\ln w_* \to +\infty.
\end{aligned}
\end{equation}
Matching thus entails that the square mass in zero field $w_0 \to w_P^+$, which, as argued above, is one of the expected hallmarks of the approach to the lower 
critical dimension. Note that in the limit process $w_0$ must remain strictly larger than the pole $w_P$ by a quantity that goes to zero with $\tilde\epsilon$: 
this again illustrates the highly singular and nonuniform approach to the lower critical dimension.

\begin{figure}
\begin{center}
\begin{picture}(230,320)
\put(0,160){\includegraphics[width=225pt]{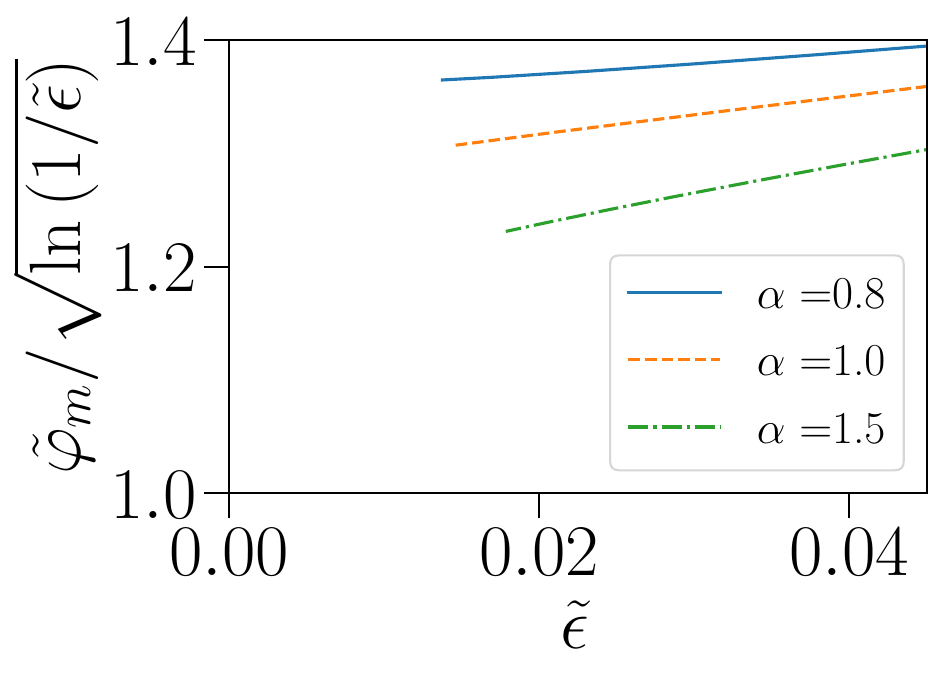}}
\put(60,215){{\huge a)}}
\put(0,0){\includegraphics[width=225pt]{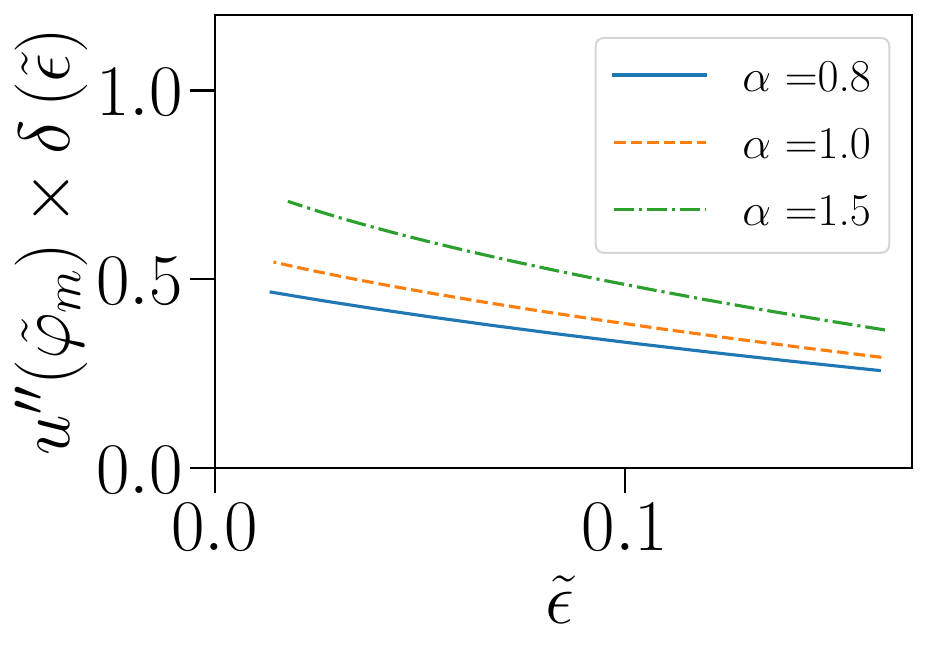}}
\put(60,60){{\huge b)}}
\end{picture}
\end{center}
\caption{(a) Location of the minimum of the potential $\varphi_{\rm m}$ divided by $\sqrt{\ln(1/\tilde\epsilon)}$ as a function of $\tilde\epsilon$.
(b) Same but for the square mass $w_{\rm m}\equiv u''(\varphi_{\rm m})$ multiplied by $\delta(\tilde\epsilon)=\tilde\epsilon\sqrt{\ln(1/\tilde\epsilon)}$. As 
predicted, both curves appear to essentially go to finite nonzero values as $\tilde\epsilon \to 0$, in agreement with our singular perturbation analysis (loglog 
corrections are not detectable with our data). On the other hand the prediction from [\onlinecite{ballhausen04}] (dashed black line in (a)) clearly does not fit 
the data. The data points in (a) and (b) are obtained from the numerical resolution of the LPA' fixed-point equations for the Exponential cutoff functions 
with several different values of the prefactor $\alpha$.}
\label{Fig_BL_scaling}
\end{figure}

To complete the proof, we note that in the chosen matching region, the leading behavior of $w(\varphi)$ in the boundary layer and that corresponding to the 
$\tilde\epsilon=0$ solution obey the same equation, $-w(\varphi) +\partial^2_\varphi[1/w(\varphi)]=0$. The difference is in the boundary condition at large field: The 
$\tilde\epsilon=0$ is limited by $\varphi_*$ while it is convenient to consider the boundary-layer one up to $\varphi_{\rm m}$. Taking this into account, the solution 
can then be obtained either as
\begin{equation}
\label{eq_matching_varphiBL}
\varphi_{\rm m}-\varphi(w)=\tilde\epsilon\varphi_{\rm m} \vert x\vert \approx \frac{\sqrt{2}}2  \frac 1{w\sqrt{\ln(\frac 1{\tilde\epsilon\varphi_{\rm m} w})}}
\end{equation}
or as
\begin{equation}
\label{eq_matching_varphi0}
\varphi_*-\varphi(w) \approx \frac{\sqrt{2}}2  \frac 1{w\sqrt{\ln(\frac {w_*}{w})}}.
\end{equation}
Matching between the two solutions is then enforced at leading order if
\begin{equation}
\begin{aligned}
\label{eq_matching}
&\varphi_*\sim\varphi_{\rm m} ,\\&
w_*\sim \frac 1{\tilde\epsilon \varphi_{\rm m}}\sim w_{\rm m},
\end{aligned}
\end{equation}
which, since $\varphi_*$ diverges as $\sqrt{\ln w_*}$ (see Eq.~(\ref{eq_phi_star}) and Appendix~\ref{app:phi_star}), immediately leads to
\begin{equation}
\begin{aligned}
\label{eq_matchingfinal}
&w_{\rm m}\sim w_* \sim\frac 1{\tilde\epsilon\sqrt{ \ln(\frac 1{\tilde\epsilon})}},\\&
\varphi_{\rm m}\sim \varphi_*\sim \sqrt{\ln(\frac 1{\tilde\epsilon})} +{\rm O}(\ln \ln (\frac 1{\tilde\epsilon})).
\end{aligned}
\end{equation}
So, as anticipated the location of the minimum of the potential $\varphi_{\rm m}$ diverges when $\tilde\epsilon\to 0$ but the width of the layer 
$\tilde\epsilon \varphi_{\rm m}$ goes to $0$. This is supported by the numerical resolution of the LPA' flow equation for values of $d$ approaching as close 
as possible the lower critical dimension: see Fig.~\ref{Fig_BL_scaling}. This result is different than the prediction of the previous FRG analysis of the 
approach to the lower critical dimension within the truncated derivative expansion in Ref.~[\onlinecite{ballhausen04}]. The latter missed the emergence of 
the boundary layer near the minimum of the potential, which led to the scaling $\varphi_{\rm m} \sim 1/\sqrt{\tilde\epsilon}$ that does not fit the data as 
shown in Fig.~\ref{Fig_BL_scaling}a.

Collecting all of the above results allows one to build a fixed-point solution $w(\varphi)\equiv u''(\varphi)$ that is valid over the whole range of 
field values when $\tilde\epsilon\to 0$. One can note the peculiar form of the present singular perturbation problem in which neither the initial condition 
in $\varphi=0$ nor the location of the layer in $\varphi_{\rm m}$ are determined {\it a priori} and must be determined through the matching procedure.

We now discuss the consequences for the LPA' prediction of the lower critical dimension $d_{\rm lc}$, the behavior of the critical temperature $T_c$, 
and the critical exponents as $\tilde\epsilon\to 0$.

\section{Results}
\label{sec:results}

\subsection{Determination of the lower critical dimension}
\label{sub:lower-critical-dim}

To determine the value of the lower critical dimension $d_{\rm lc}$ we consider the last of the LPA' equations that we have not yet used, {\it i.e.}, 
Eq.~(\ref{eq_anomalous-dim}) for the anomalous dimension of the field. This equation involves the square mass $w(\varphi)$ in the boundary layer only. 
When $\tilde\epsilon\to 0$, $w(\varphi_{\rm m})\to \infty$, $\eta\to 2-d$, and the threshold function $m_{4,0}^{(d)}(w;\eta)$ can be replaced in 
Eq.~(\ref{eq_anomalous-dim}) by its asymptotic form,
\begin{equation}
\begin{aligned}
\label{eq_threshold_m40_asympt}
m_{4,0}^{(d)}(w;2-d)\sim \frac {\alpha d A_d-\alpha^2 B_d}{w^4}
\end{aligned}
\end{equation}
where $A_d$ is given in Eq.~(\ref{eq_Ad}) and, as derived in Appendix~\ref{app:threshold},
\begin{equation}
\begin{aligned}
\label{eq_Bd}
B_d=\frac{3d-2}{2} \int_0^{\infty}dy y^{\frac d2} [(y\frac{r(y)}\alpha)']^2\,.
\end{aligned}
\end{equation}
As before a prime indicates a derivative with respect to the argument of the function and the IR cutoff functions that we use are defined in 
Eq.~(\ref{eq_cutoff_choices}).

With the rescaling of the field and the notations introduced in Sec.~\ref{sub:inner-solutionBL}, one can then rewrite Eq.~(\ref{eq_anomalous-dim}) as
\begin{equation}
\label{eq_anomalous-dim_lc}
2-d = 2(d-\alpha \,\frac{B_d}{A_d}) \frac{g'(0)^2}{g(0)^4}=\frac{4}{\pi}(d-\alpha\, \frac{B_d}{A_d})
\end{equation}
where we have used that the solution within the layer around the minimum satisfies Eqs.~(\ref{eq_BLg_min}) and (\ref{eq_g0min}). The solution of 
the above equation gives $d=d_{\rm lc}$.

\begin{figure}
\includegraphics[width=0.8\linewidth]{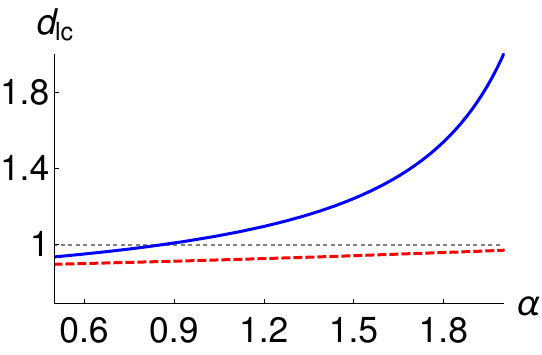}
\caption{Lower critical dimension $d_{\rm lc}(\alpha)$ predicted by the solution of the LPA' with the Theta (upper blue curve) and Exponential 
(lower red curve) IR cutoff functions.}
\label{Fig_d_lc}
\end{figure}

For the Theta cutoff function, $B_d/A_d=(3d-2)/4$ (see Appendix~\ref{app:threshold}), so that we obtain an explicit analytical expression for the lower 
critical dimension:
\begin{equation}
\label{eq_lc_theta}
d_{\rm lc}(\alpha)=2 \frac{\pi-\alpha}{\pi +4-3\alpha},
\end{equation}
which for instance predicts $d_{\rm lc}=1.03419\cdots$ for $\alpha=1$. Note that the solution derived in the previous section required $d<2$. This 
entails that $\alpha<2$, so that the pole in $\alpha=(\pi+4)/3$ is not attained. The variation of $d_{\rm lc}$ with $\alpha$ is shown in 
Fig.~\ref{Fig_d_lc}.

For the Exponential cutoff function, one finds that $B_d/A_d=2^{-(1+d/2)}(3d-2)/4$ (see Appendix~\ref{app:threshold}), so that $d_{\rm lc}$ is solution 
of the implicit equation
\begin{equation}
\label{eq_lc_theta}
d_{\rm lc}(\alpha)=2 \frac{\pi-\alpha_*(d)}{\pi +4-3\alpha_*(d)}
\end{equation}
with $\alpha_*(d)= 2^{-(1+d/2)}\alpha$. The outcome is plotted in Fig.~\ref{Fig_d_lc}.

We therefore obtain that for a reasonable range of the variational parameter $\alpha$ (which cannot be fixed by any principle of minimum sensitivity on the 
variation of $d_{\rm lc}(\alpha)$ because the latter does not display a minimum at the present level of approximation) the predicted lower critical dimension 
is indeed close to the exact result, $d_{\rm lc}=1$ (within $10\%$ for the Exponential cutoff function; the result found with the Theta regulator appears less 
well-behaved than that of the Exponential cutoff).

\subsection{Critical temperature $T_c$ as $\tilde\epsilon\to 0$}
\label{sub:Tc}

One of the many defining properties of the lower critical dimension is that the critical temperature $T_c$ goes to zero. This is a bare quantity which is 
not easily retrieved from the RG flow. However when it goes to zero, a simple reasoning based on the Boltzmann form of the distribution suggests that 
the field scales as the square root of the inverse temperature. As the dimension of the field at criticality goes to zero at the lower critical dimension, one therefore 
expects that $T_c\sim 1/\varphi_{\rm m}^2$. This is indeed what is found in the correspondance between the Wilsonian dimensionless action of the O($N>2$) 
model and the nonlinear sigma model near the lower critical dimension $d=2$.\cite{???} Together with Eq.~(\ref{eq_matchingfinal}), this scaling leads to
\begin{equation}
\label{eq_Tc}
T_c\sim \frac 1{\ln(\frac 1{\tilde\epsilon})}\to 0
\end{equation}
when $\tilde\epsilon\to 0$. 

Recast in terms of the field dimension $D_\phi=(d-2+\eta)/2$ the above expression is equivalent to
\begin{equation}
\label{eq_Tc_bis}
T_c\propto \frac 1{\ln(\frac 1{D_\phi})}.
\end{equation}
This relation is similar to that obtained by Bruce and Wallace from a detailed droplet theory.\cite{bruce81,bruce83} In the latter, the expansion is 
performed in $\epsilon=d-d_{\rm lc}$ with $d_{\rm lc}=1$. The outcome is that $T_c$ has a simple expansion in powers of $\epsilon$, 
$T_c\propto \epsilon + {\rm O}(\epsilon^2)$, but $D_\phi$ has instead a singular behavior, with $D_\phi \sim e^{-2/\epsilon}/\epsilon$. Combining 
the two gives Eq.~(\ref{eq_Tc_bis}). Note that this relation is not verified by the prediction of Ref.~[\onlinecite{ballhausen04}].

\subsection{Stability of the fixed point, essential scaling, and correlation-length critical exponent as $\tilde\epsilon\to 0$}
\label{sub:stability}

The stability of the fixed point can be studied by looking at perturbations around it and the resulting eigenvalue equation equation obtained in linear order of 
the perturbation. For the present LPA' approximation, after introducing small perturbations around the fixed point as $w_k(\varphi)=w(\varphi)+
k^{\lambda} \delta w(\varphi)$, $\eta_k=\eta+ k^{\lambda} \delta \eta$, $\varphi_{\rm m k}=\varphi_{\rm m}+k^\lambda \delta\varphi_{\rm m}$, etc., 
with $\lambda$ an eigenvalue to be determined, the linearized equation for $\delta w(\varphi)$ reads
\begin{equation}
\begin{aligned}
\label{eq_eigenvalue}
&\lambda \delta w(\varphi)= -(2-\eta) \delta w(\varphi) +(2-\eta)\tilde\epsilon\varphi \delta w'(\varphi) + \\&
2 v_d \partial^2_\varphi [\partial_w\ell_0^{(d)}(w(\varphi);\eta)\delta w(\varphi)]+ \big (w(\varphi)+\frac 12 \varphi w'(\varphi) \\&
+ 2 v_d \partial^2_\varphi [\partial_\eta\ell_0^{(d)}(w(\varphi);\eta)]\big )\delta\eta ,
\end{aligned}
\end{equation}
and the expressions for $\delta\eta$ and $\delta\varphi_{\rm m}$ are given in Appendix~\ref{app:stability}. We are especially interested in finding the relevant 
eigenvalue that gives the correlation length exponent $\nu$ which is known to diverge at the lower critical dimension in an exact treatment.

As we did for the fixed point, one can attempt a singular perturbation analysis when $\tilde\epsilon\to 0$ (and $\eta\to 2-d$) by looking separately at the 
$\tilde\epsilon= 0$ equation for $\varphi$ of O(1) and at an equation in terms of the scaled variable $x=(\varphi-\varphi_{\rm m})/(\tilde\epsilon\varphi_{\rm m})$ 
near the minimum $\varphi_{\rm m}$. However, one immediately sees that if $\lambda={\rm O}(\tilde\epsilon)$ or more generally goes to zero when 
$\tilde\epsilon\to 0$, which is the expected behavior of the relevant eigenvalue(s), working at the leading order in $\tilde\epsilon$ does not allow the 
determination of $\lambda$ beyond the fact that it starts as $0$. 

This can be illustrated by considering one eigenvalue that can be exactly obtained together with its eigenfunction. One easily finds that 
$\lambda=-(2-\eta)\tilde\epsilon$ is a solution of Eq.~(\ref{eq_eigenvalue}) with
\begin{equation}
\begin{aligned}
\label{eq_eigenperturb_trivial}
&\delta w(\varphi)= \delta K \, w'(\varphi)\\&
\delta \eta=0=\delta \tilde\epsilon \\&
\delta \varphi_{{\rm m}}=-\delta K
\end{aligned}
\end{equation}
with $\delta K$ a constant that can be taken as infinitesimal to linearize the RG flow equations. Note that despite the fact that it corresponds to a relevant 
direction around the fixed point this eigenvalue is not the one we are interested in because it is associated with an odd ($Z_2$ antisymmetric) perturbation. We 
would instead like to determine the relevant eigenvalue associated with an even ($Z_2$ symmetric) perturbation which gives the correlation-length exponent $\nu$ 
through $1/\nu=-\lambda$. It is nonetheless instructive to study how the exact result for $\lambda=-(2-\eta)\tilde\epsilon$ translates into the leading order of 
the singular perturbation analysis and we trivially find that only $\lambda=0$ can be obtained by working at the leading order of the $\tilde\epsilon=0$ and of 
the boundary-layer equations.

\begin{figure}
\begin{center}
\begin{picture}(230,320)
\put(0,160){\includegraphics[width=225pt]{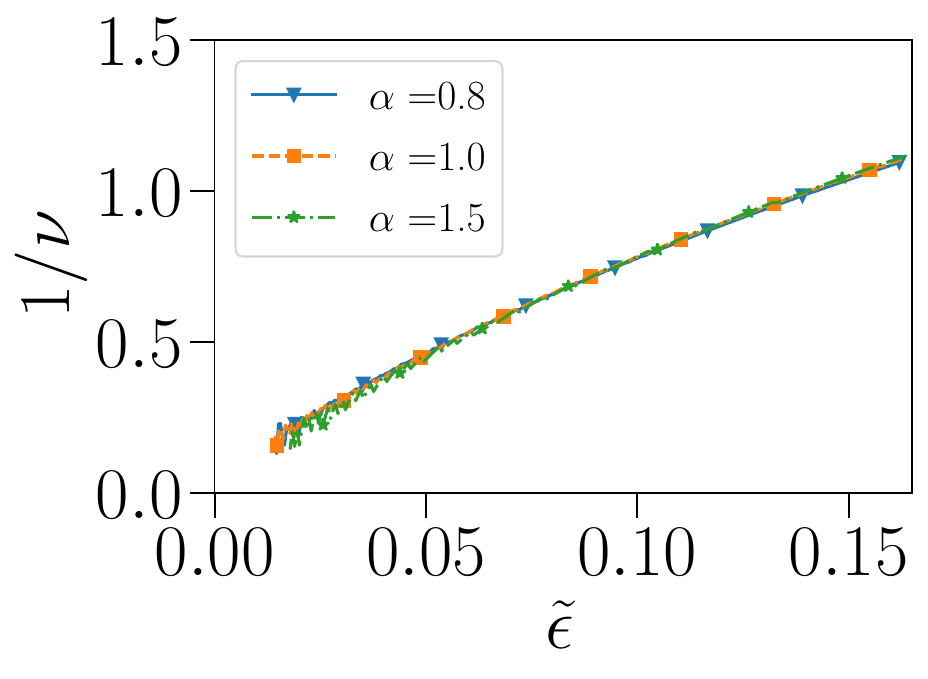}}
\put(190,215){{\huge a)}}
\put(0,0){\includegraphics[width=225pt]{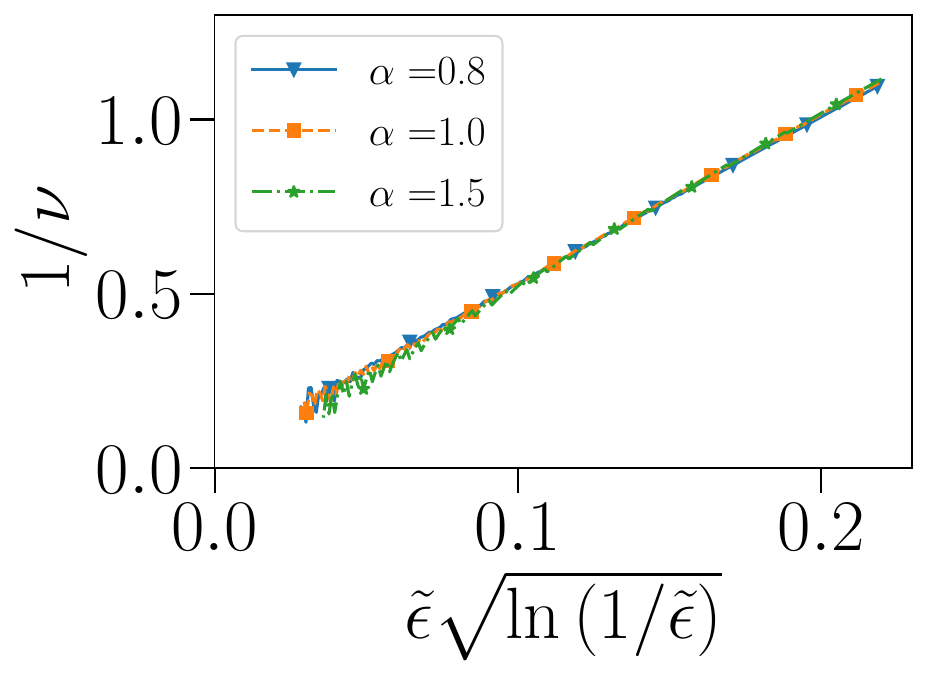}}
\put(190,60){{\huge b)}}
\end{picture}
\end{center}
\caption{The relevant eigenvalue $1/\nu$ obtained from the numerical resolution of the LPA' equations with the Exponential IR cutoff function and several values of 
the parameter $\alpha$. It is plotted as a function of $\tilde\epsilon$ (a) and $\tilde\epsilon \sqrt{\ln(1/\tilde\epsilon)}\sim \delta(\tilde\epsilon)$ (b).}
\label{Fig_eigenvalues}
\end{figure}

This example confirms that eigenvalues going to zero as $\tilde\epsilon\to 0$ cannot be determined from the singular perturbation analysis at the leading order. 
One needs to go to the next order. In the present case this seems a formidable task that we will not undertake. We instead perform a numerical 
investigation by solving the LPA' eigenvalue equation, Eq.~(\ref{eq_eigenvalue}), together with the fixed-point equation at fixed $d$, trying to reach as low as 
possible values near the lower critical dimension. As it should, we find that the critical fixed point has two relevant eigendirections: one corresponds to an even 
eigenfunction and gives the critical exponent $\nu$ and the other is equal to $-(2-\eta)\tilde\epsilon=-D_\phi$ and is associated with an odd eigenfunction related 
to the magnetic field (the scaling dimension of the magnetic field is then $d-D_\varphi$). All the other eigenvalues are positive, {\it i.e.}, irrelevant, and of O(1) 
when $\tilde\epsilon\to 0$, as it should for a critical fixed point.  We show $1/\nu$ as obtained for the Exponential IR cutoff function and several values of the 
parameter $\alpha$  in Fig.~\ref{Fig_eigenvalues}. It is plotted both versus $\tilde\epsilon$ and versus the boundary-layer width 
$\delta(\tilde\epsilon)\sim \tilde\epsilon \sqrt{\ln(1/\tilde\epsilon)}$. We observe that $1/\nu$ seems to be heading toward $0$ when $\tilde\epsilon \to 0$, which is 
the expected behavior for an essential scaling of the correlation length as one approaches the lower critical dimension. Over the accessible range of dimensions, it  
appears to do so slower than linearly in $\tilde\epsilon$, possibly like $\tilde\epsilon \sqrt{\ln(1/\tilde\epsilon)}$. On the other hand the behavior is not compatible 
with the prediction of the droplet theory which would be $1/\ln(1/\tilde\epsilon)$.\cite{bruce81,bruce83,wallace84} However, our 
numerical results may not yet be in the asymptotic regime near $d_{\rm lc}$ and the conclusion should therefore be taken with a grain of salt.

\section{Conclusion}
\label{sec_conclusion}

We have presented a functional renormalization group (FRG) description of the approach to the lower critical dimension $d_{\rm lc}$ in a scalar $\varphi^4$ 
theory by using one of the simplest nonperturbative approximation level obtained as a truncation of the derivative expansion, the so-called LPA'. Our purpose 
is to test how a generic approximation scheme that works across dimensions in a continuous way and has been shown to be accurate in dimensions 
$d\geq 2$ for instance\cite{dupuis21} is able to describe dimensions close to the lower critical dimension in a system with a discrete symmetry where 
it is known that the long-distance physics is controlled by the proliferation of localized excitations (in the present case, droplets that become point-like kinks and 
anti-kinks at the lower critical dimension $d_{\rm lc}=1$\cite{bruce81,bruce83,wallace84}). We show that the limit of $d$ going to $d_{\rm lc}$ for the fixed-point 
effective action is nonuniform in the (average) field, with the emergence of a boundary (or more precisely, interior) layer around the minimum of the 
dimensionless potential. The minimum goes to infinity and the width of the layer goes to zero as $d\to d_{\rm lc}$, at odds with the outcome of an earlier FRG 
study.\cite{ballhausen04} The behavior of the critical temperature $T_c$ is compatible with the expected exact results and, although the prediction of $d_{\rm lc}$ is 
dependent on the infrared regulator used in the FRG, we find it rather close to the exact value $d_{\rm lc}=1$ (within $10\%$ for a range of regulators).

The next step is to check that the scenario found at the LPA' is valid at all orders of the truncated derivative expansion. Work is now in progress to 
investigate the next order, which includes a field-renormalization function in addition to the potential. Preliminary results indicate that the same mechanism 
of a nonuniform convergence to the lower critical dimension with the emergence of an interior layer around the minimum of the dimensionless potential is 
also at play. The next order of the derivative expansion also seems to more properly describe the form of divergence of the correlation length exponent $\nu$.

As already stressed, our goal is not to provide yet another theoretical description of the approach to the lower critical dimension for systems in the universality 
class of the Ising model, a question which has been quite well understood for several decades. It is to benchmark a generic nonperturbative but approximate 
FRG approach to later tackle problems that are still open such as the value of the lower critical dimension of the athermally driven random-field Ising model 
(RFIM).\cite{sethna} The lower critical dimension of the RFIM in equilibrium has been rigorously shown to be $d_{\rm lc}=2$\cite{aizenman-wehr}, but that 
for the far-from-equilibrium driven RFIM is debated.\cite{shukla,spasojevic,hayden-sethna} Finally one might also hope that the present solution near the lower 
critical dimension can suggest new approximation schemes of the FRG that are not necessarily based on the truncation of the derivative expansion.

\acknowledgements
We thank Adam Ran\c con for numerous discussions related to this topic during the years. IB wishes to thank Nicolas Wschebor and Maroje Marohni\'c for useful discussions and feedbacks. LNF and IB acknowledge the support of the the QuantiXLie Centre of Excellence, a project cofinanced by the Croatian Government and European Union through the European Regional Development Fund - the Competitiveness and Cohesion Operational Programme (Grant KK.01.1.1.01.0004).

\appendix

\section{FRG flow equations}
\label{appx_flows}

The beta functional describing the FRG flow of the dimensionless effective potential $u_k(\varphi)$ is given by\cite{berges02}
\begin{equation}
\begin{aligned}
\label{eq_beta_potential}
&\beta_u(\varphi;\eta)=2 v_d \ell_0^{(d)}(u''(\varphi);\eta,z(\varphi)),
\end{aligned}
\end{equation}
where $v_d^{-1}=2^{d+1} \pi^{d/2}\Gamma(d/2)$; $\ell_n^{(d)}$ is a (strictly positive) dimensionless threshold function defined by
\begin{equation}
\begin{aligned}
\label{eq_threshold_ell}
\ell_n^{(d)}(w;\eta,z)= -\left (\frac{n+\delta_{n,0}}2\right )\int_0^{\infty}dy y^{\frac d2}\frac{\eta r(y)+2yr'(y)}{(y[z+r(y)]+w)^{n+1}}
\end{aligned}
\end{equation}
where the dimensionless infrared cutoff function (or IR regulator) $r(y)$ is obtained from the dimensionful one, $R_k(q^2)$, 
introduced in Eq.~(\ref{eq_regulator}) through 
\begin{equation}
\label{eq_IRregulator}
R_k(q^2)=Z_k k^2 y\, r(y) \;\;\; {\rm with}\;\; y=\frac {q^2}{k^2}
\end{equation}
with $k$ is the running IR cutoff and $Z_k$ the dimensionful field renormalization (such that the running anomalous dimension is defined by 
$\eta_k=-k\partial_k Z_k$).

From the exact FRG equation for the 2-point 1-PI correlation function evaluated for a uniform field configuration one can extract the beta functional for the 
dimensionless field renormalization function $z_k(\varphi)$,\cite{berges02}
\begin{equation}
\begin{aligned}
\label{eq_beta_fieldrenormalization}
&\beta_z(\varphi;\eta)= - \frac{4v_d}d u'''(\varphi)^2m_{4,0}^{(d)}(u''(\varphi);\eta,z(\varphi)) - \\&
\frac {8v_d}d u'''(\varphi)z'(\varphi)m_{4,0}^{(d+2)}(u''(\varphi);\eta,z(\varphi))\\&
- \frac{4v_d}d z'(\varphi)^2 m_{4,0}^{(d+4)}(u''(\varphi);\eta,z(\varphi)) - 2v_d z''(\varphi) \times \\&
\ell_1^{(d)}(u''(\varphi);\eta,z(\varphi))+  4 v_d u'''(\varphi) z'(\varphi) \ell_2^{(d)}(u''(\varphi);\eta,z(\varphi)) \\&
+ 2 v_d \frac{1+2d}d z'(\varphi)^2 \ell_2^{(d+2)}(u''(\varphi);\eta,z(\varphi)),
\end{aligned}
\end{equation}
where $m_{n,0}^{(d)}$ is another (strictly positive) threshold function defined as
\begin{equation}
\begin{aligned}
\label{eq_threshold_m}
&m_{n,0}^{(d)}(w;\eta,z)= \frac 12 \int_0^{\infty}dy y^{\frac d2} \frac{z+(yr(y))'}{(y[z+r(y)]+w)^n}
\bigg [2\eta (yr(y))'\\&
+4(y^2r'(y))' -n \,\frac {y[z+(yr(y))'][\eta r(y)+2yr'(y)]}{y[z+r(y)]+w}  \bigg ].
\end{aligned}
\end{equation}
To derive Eq.~(\ref{eq_beta_fieldrenormalization}) we have neglected the higher-order terms in Eq.~(\ref{eq_DE}) which involve four spatial derivatives: It therefore 
represents the second order of the derivative expansion which is fully characterized by the two functions $U_k(\phi)$ and $Z_k(\phi)$. Three more 
functions are required at the order O($\partial^4$), etc.

\section{Properties of the threshold functions}
\label{app:threshold}

The threshold functions introduced in the main text and in the above Appendix are a strictly positive dimensionless functions that enforce the decoupling 
of the low-momentum and high-momentum modes.\cite{berges02} We only consider the LPA' approximation so that the dimensionless field renormalization 
function $z(\varphi)\equiv 1$, but this is easily generalizable.

Before discussing some of their generic properties it is illustrative to give their explicit expression for a specific choice of IR cutoff function, 
$r(y)=\Theta(1-y)(1-y)/y$, which is the Theta cutoff function with $\alpha=1$ (also called Litim or optimized regulator\cite{litim01}):
\begin{equation}
\begin{aligned}
\label{eq_litim}
&\ell_{n}^{(d)}(w;\eta)=\frac {2(d+2-\eta)}{d (d+2)}\frac n{(1+w)^{n+1}}\\&
m_{n,0}^{(d)}(w;\eta)=\frac 1{(1+w)^{n}}.
\end{aligned}
\end{equation}
One can see that the threshold functions monotonically decrease as $w$ increases, blow up near the pole of the propagator $w_P$ (here $w_P=-1$) and go to 
zero as power laws when $w\to +\infty$. They are defined for $w>w_P$.

We analyze the behavior of the threshold functions in two limiting cases, when the mass $w$ is large and when it approaches the pole of the propagator, 
{\it i.e.}, when $w+\min_y{\{y(1+r(y))\}} \to 0$.

When the mass $w\to \infty$ and $z=1$  one easily finds from Eq.~(\ref{eq_threshold_ell}) that
\begin{equation}
\begin{aligned}
\ell_n^{(d)}(w;\eta)\sim \frac{ A_n^{(d)}(\eta)}{w^{n+1}} + {\rm O}(\frac 1{w^{n+2}})
\end{aligned}
\end{equation}
where
\begin{equation}
\begin{aligned}
A_n^{(d)}(\eta)&= -\left (\frac{n+\delta_{n,0}}2\right ) \int_0^{\infty}dy y^{\frac d2}[\eta r(y)+2yr'(y)]\\&=(n+\delta_{n,0})A_0^{(d)}(\eta)
\end{aligned}
\end{equation}
and $A_0^{(d)}(\eta)$ can be rewritten as
\begin{equation}
\begin{aligned}
\label{eq_coeffA}
A_0^{(d)}(\eta)= \frac{d+2-\eta}2 \int_0^{\infty}dy y^{\frac d2}r(y).
\end{aligned}
\end{equation}
The choices of $r(y)$ that we use in this work are given in Eq.~(\ref{eq_cutoff_choices}) so that $A_0^{(d)}(\eta)$ is proportional to $\alpha$. When 
$\eta=2-d$ this leads to the expression of $A_d$ in Eq.~(\ref{eq_Ad}).

Similarly, from Eq.~(\ref{eq_threshold_m}) one finds
\begin{equation}
\begin{aligned}
m_{n,0}^{(d)}(w;\eta)\sim \frac{-B^{(d)}(\eta)+d A_0^{(d)}(\eta)}{w^n} +{\rm O}(\frac 1{w^{n+1}})\,,
\end{aligned}
\end{equation}
with
\begin{equation}
\begin{aligned}
\label{eq_coeffB}
&B^{(d)}(\eta)=  -\int_0^{\infty}dy y^{\frac d2}(yr (y))'[\eta (yr(y))'+2(y^2r'(y))']\\&
= \frac{d+2-2\eta}2 \int_0^{\infty}dy y^{\frac d2}[(yr (y))']^2.
\end{aligned}
\end{equation}
When $\eta=2-d$ one immediately obtains Eq.~(\ref{eq_Bd}).

All of the above results of course match with the expansion of the expressions in Eq.~(\ref{eq_litim}).

We now turn to the expression of the threshold functions near the pole of the propagator, when $w\to w_P=-\min_y{\{y(1+r(y))\}}$. Note that the FRG equations 
are well behaved for $w+\min_y{\{y(1+r(y))\}}>0$. The approach to the pole is what controls the return to convexity of the effective potential in the ordered 
phase\cite{tetradis92,pelaez-wschebor,berges02} and is therefore important in the vicinity of the lower critical dimension where the critical fixed and the fixed 
point describing the ordered phase merge.

For the Theta cutoff function and for $z=1$, 
\begin{equation}
\begin{aligned}
\ell_{n}^{(d)}(w;\eta)=\alpha \frac{n+\delta_{n,0}}2 \int_0^1dy y^{\frac d2-1}\frac {[(2-\eta)+\eta y]}{[w+\alpha+(1-\alpha)y]^{n+1}}
\end{aligned}
\end{equation}
\begin{equation}
\begin{aligned}
&m_{n,0}^{(d)}(w;\eta)=\\&
\alpha\frac{2-\alpha}{[w+1]^n}-\alpha\eta(1-\alpha)\int_0^1dy y^{\frac d2}\frac 1{[w+\alpha+(1-\alpha)y]^n}+\\&
\alpha \frac{n}2 (1-\alpha)^2\int_0^1dy y^{\frac d2} \frac{[(2-\eta)+\eta y]}{[w+\alpha+(1-\alpha)y]^{n+1}}\,.
\end{aligned}
\end{equation}

The pole of the propagator $1/[w+\alpha+(1-\alpha)y]$ is $w_P=-\alpha$  if $\alpha< 1$, and is reached in $y=0$, and is $w_P=-1$ if $\alpha>1$, and is 
reached in $y=1$. The case $\alpha=1$ corresponds to the expressions in Eq.~(\ref{eq_litim}) and the approach to the pole in $w_P=-1$ can be read off directly. The 
threshold functions generically diverge as inverse power laws of $(w+w_P)$ when $w$ approaches the pole $w_P$ which is either $-\alpha$ or $-1$. An exception 
is $\ell_{0}^{(d)}(w;\eta)$ which behaves as $-\ln(w+1)$ when $\alpha>1$.

For the Exponential cutoff function (and for $z=1$), a similar behavior is encountered except that the pole is attained either in 
$y=0$ and is equal to $w_P=-\alpha$ when $\alpha<1$ or in $y=\ln\alpha$ and is equal to $w=-(1+\ln\alpha)$ when $\alpha>1$. The divergence of the threshold 
function as $w+w_P\to 0^+$ is generically power-law-like.

\section{Further analysis of the $\tilde\epsilon=0$ solution}
\label{app:phi_star}

To prove Eq.~(\ref{eq_phi_star}) we start from Eq.~(\ref{eq_epsilon-zero3}) where we recall that $F(\Phi)\equiv w(\Phi)$ and $\Phi(w)=\ell_0^{(d)}(w;2-d)$. From 
the analysis of the threshold functions in the preceding section, one can infer that $w(\Phi)$ is a monotonically decreasing function that starts from 
$+\infty$ when $\Phi=0$ and asymptotically goes to the pole $w_P<0$ when $\Phi\to +\infty$. In the regime of interest where $w_0=w(0)\to w_P^+$ and 
$w_*=w(\varphi_*)\to +\infty$, the relevant range of $\Phi$ is from $\Phi_*\sim\alpha A_d/w_*\to 0$ (see Eq.~(\ref{eq_ell0-asympt}) to 
$\Phi_0 \sim \alpha (A_d/\vert w_P\vert)\ln w_*\to +\infty$ (see Eq.~(\ref{eq_w*3}). 

Eq.~(\ref{eq_epsilon-zero3}) can be rewritten as
\begin{equation}
\label{eq_varphi_Phi1}
\frac{\partial\Phi}{\partial\varphi}=-\sqrt{\frac d{v_d}} \sqrt{-\int_{\Phi}^{\Phi_0} d\Phi'w(\Phi')},
\end{equation}
with $\Phi(\varphi)$ is a monotonically decreasing function between $0$ and $\varphi_*$. (Note that by definition $\int_{\Phi_*}^{\Phi_0} d\Phi \,w(\Phi )=0$.) 
This leads to
\begin{equation}
\label{eq_varphi_Phi2}
\varphi(\Phi)=\sqrt{\frac{v_d}d}\int_{\Phi}^{\Phi_0} \frac{d\Phi'}{\sqrt{-\int_{\Phi'}^{\Phi_0} d\Phi'' w(\Phi'')}}\,,
\end{equation}

We define $\Phi_{\rm i}$ and the associated field $\varphi_{\rm i}$ such that $w(\Phi_{\rm i})=w(\varphi_{\rm i})=0$. Then,
\begin{equation}
\label{eq_varphi_i}
\varphi_{\rm i}=\sqrt{\frac{v_d}d} \int_{\Phi_{\rm i}}^{\Phi_0} \frac{d\Phi}{\sqrt{-\int_{\Phi}^{\Phi_0} d\Phi'w(\Phi')}}\,,
\end{equation}
where $\Phi_{\rm i}=\ell_0^{(d)}(w=0)$ is of O($1$) and $w(\Phi)$ is monotonically decreasing and negative in the interval between $\Phi_{\rm i}$ and $\Phi_0$. 
From the properties of the threshold functions it is easily check that that $w(\Phi)$ is concave: indeed, its second derivative is 
$w''(\Phi)=-\partial_w^2\ell_0^{(d)}(w)/[\partial_w\ell_0^{(d)}(w)]^3$ with $\partial_w\ell_0^{(d)}(w)=-\ell_1^{(d)}(w)<0$ and 
$\partial_w^2\ell_0^{(d)}(w)=\ell_2^{(d)}(w)>0$. In consequence, when $\Phi_{\rm i}\leq \Phi\leq \Phi_0$,
\begin{equation}
\vert w_0\vert \geq -w(\Phi)\geq \vert w_0\vert \frac{\Phi-\Phi_{\rm i}}{\Phi_0-\Phi_{\rm i}}.
\end{equation}
When inserted in Eq.~(\ref{eq_varphi_i}), after some elementary algebra and using the asymptotic behavior of $\Phi_0$ when $w_*\to \infty$, this implies that
\begin{equation}
\label{eq_varphi_i_2}
\frac{\sqrt{2}}2\pi\sqrt{\frac{\alpha v_dA_d}{d \,w_P^2}}\sqrt{\ln w_*}\geq \varphi_{\rm i}\geq 2\sqrt{\frac{\alpha v_dA_d}{d\, w_P^2}}\sqrt{\ln w_*}\,,
\end{equation}
which proves that $\varphi_{\rm i}\sim \sqrt{\ln w_*}$.

To complete the demonstration for all $\varphi$'s between $\varphi_{\rm i}$ and $\varphi_{*}$ we can rewrite Eqs.~(\ref{eq_varphi_Phi2}) and (\ref{eq_varphi_i}) 
as
\begin{equation}
\begin{aligned}
\label{eq_varphi_*}
\varphi_{*}-\varphi_{\rm i}=\sqrt{\frac{v_d}d} \int_{\Phi_*}^{\Phi_{\rm i}} \frac{d\Phi}{\sqrt{\int_{\Phi_*}^{\Phi} d\Phi'w(\Phi')}} \,.
\end{aligned}
\end{equation}
By using the properties of the function $w(\Phi)$ we find that $\int_{\Phi_*}^{\Phi} d\Phi'w(\Phi')$ is a monotonically increasing and concave function of $\Phi$ for 
$\Phi \leq \Phi_{\rm i}$, which implies that
\begin{equation}
\begin{aligned}
\label{eq_bound}
\int_{\Phi_*}^{\Phi} d\Phi'w(\Phi') \geq \int_{\Phi_*}^{\Phi_i} d\Phi w(\Phi) \left (\frac{\Phi-\Phi_*}{\Phi_i-\Phi_*}\right ).
\end{aligned}
\end{equation}
Then,
\begin{equation}
\begin{aligned}
\label{eq_varphi_*2}
\varphi_{*}-\varphi_{\rm i}&\leq \sqrt{\frac{v_d}d} \sqrt{\frac{\Phi_i-\Phi_*}{\int_{\Phi_*}^{\Phi_i} d\Phi w(\Phi)}}\int_{\Phi_*}^{\Phi_{\rm i}} \frac{d\Phi}{\sqrt{\Phi-\Phi_*}}\\&
\leq 2 \sqrt{\frac{v_d}d} \frac{\Phi_i-\Phi_*}{\sqrt{\int_{\Phi_*}^{\Phi_i} d\Phi w(\Phi)}},
\end{aligned}
\end{equation}
where we recall that $\Phi_{\rm i}=\ell_0^{(d)}(w=0)={\rm O}(1)$ and $\Phi_*\sim \alpha A_d/w_* \to 0$. After rewriting 
$\int_{\Phi_*}^{\Phi_i} d\Phi w(\Phi)=-\int_0^{w_*}dw w \partial_w\ell_0^{(d)}(w)$, integrating by part and using the properties of the threshold function 
$\ell_0^{(d)}$, one obtains that the integral behaves as $\alpha A_d \ln w_*$ when $w_* \gg 1$. This finally leads to
\begin{equation}
\begin{aligned}
\label{eq_varphi_*3}
\varphi_{*}-\varphi_{\rm i}\lesssim 2 \sqrt{\frac{v_d}{\alpha d A_d}} \frac{\ell_0^{(d)}(w=0)}{\sqrt{\ln w_*}}\,,
\end{aligned}
\end{equation}
so that $\varphi_*\sim \varphi_{\rm i}\sim \sqrt{\ln w_*}$, as announced.

\section{Eigenvalue equations}
\label{app:stability}

the linearized equation for the perturbation of the square-mass function $k^\lambda \delta w(\varphi)$ in Eq.~(\ref{eq_eigenvalue}) should be complemented by 
linearized equations for the perturbation of the anomalous dimension $\delta\eta$ and of the minimum of the potential $\delta\varphi_{\rm m}$. That for $\delta\eta$ 
follows directly from Eq.~(\ref{eq_anomalous-dim}) which is valid for all scales $k$ in the LPA'. It reads
\begin{equation}
\begin{aligned}
\label{eq_anomalous-dim_linear}
\delta\eta=& \frac{4v_d}d \Big (2w'(\varphi_{\rm m}) m_{4,0}^{(d)}(w(\varphi_{\rm m});\eta)[\delta w'(\varphi_{\rm m})+w''(\varphi_{\rm m})\delta\varphi_{\rm m}] \\&
+w'(\varphi_{\rm m})^2 \partial_w m_{4,0}^{(d)}(w(\varphi_{\rm m});\eta)[\delta w(\varphi_{\rm m})+w'(\varphi_{\rm m})\delta\varphi_{\rm m}] \\&
+ w'(\varphi_{\rm m})^2 \partial_\eta m_{4,0}^{(d)}(w(\varphi_{\rm m});\eta)\delta\eta \Big )
\end{aligned}
\end{equation}
and does not involve $\lambda$ explicitly.

We next need the flow equation for the $k$-dependent minimum which is obtained from that of $u'_k(\varphi)$ as
\begin{equation}
\begin{aligned}
\label{eq_phim_flow}
\partial_t \varphi_{{\rm m}k}=&-\frac{(d-2+\eta_k)}{2}\varphi_{{\rm m},k}-2v_d\,\frac{w_k'(\varphi_{{\rm m}k})}{w_k(\varphi_{{\rm m}k})} \times\\&
\partial_w\ell_0^{(d)}(w;\eta_k)\vert_{w=w_k(\varphi_{{\rm m}k})}.
\end{aligned}
\end{equation}
Linearizing then leads to
\begin{equation}
\begin{aligned}
\label{eq_phim_flow}
&\lambda \delta\varphi_{{\rm m}}=-\Big [\frac{(d-2+\eta)}{2} +2v_d\ell_1^{(d)}(w(\varphi_{{\rm m}});\eta)
\Big (\frac{w'(\varphi_{{\rm m}})^2}{w(\varphi_{{\rm m}})^2}- \\&
\frac{w''(\varphi_{{\rm m}})}{w(\varphi_{{\rm m}})}\Big )-4v_d \ell_2^{(d)}(w(\varphi_{{\rm m}});\eta)\frac{w'(\varphi_{{\rm m}})^2}{w(\varphi_{{\rm m}})}\Big ] 
\delta\varphi_{{\rm m}}+ \\&
2v_d \Big [\ell_1^{(d)}(w(\varphi_{{\rm m}});\eta) \Big (\frac{\delta w'(\varphi_{{\rm m}})}{w(\varphi_{{\rm m}})}-
\frac{w'(\varphi_{{\rm m}})\delta w(\varphi_{{\rm m}})}{w(\varphi_{{\rm m}})^2}\Big ) + \\&
2 \ell_2^{(d)}(w(\varphi_{{\rm m}});\eta)\frac{w'(\varphi_{{\rm m}})\delta w(\varphi_{{\rm m}})}{w(\varphi_{{\rm m}})}\Big ] -\delta\eta \Big [\frac{\varphi_{{\rm m}}}2 + \\&
2 v_d \partial_\eta \ell_1^{(d)}(w(\varphi_{{\rm m}});\eta)\frac{w'(\varphi_{{\rm m}})}{w(\varphi_{{\rm m}})}\Big ]
\end{aligned}
\end{equation}
where we have used the property of the threshold functions that
$\partial_w\ell_n^{(d)}(w;\eta)=-(n+1)\ell_{n+1}^{(d)}(w;\eta)$.
\\


\begin{thebibliography}{45}
\bibitem{footnote_2d}
A similar difficulty for the truncated derivative expansion of the FRG arises for the 2-dimensional linear O(2) model when trying to capture the asymptotic 
critical BKT behavior\cite{berezinskii,KT} which is controlled by localized vortices.\cite{wetterich-XY,jakubczyk-XY} On the other hand, the 
instantons/solitons in the 2-dimensional sine Gordon model are actually extended defects in one dimension (and localized kinks or anti-kinks in the 
other) and are well described by low orders of the derivative expansion.\cite{daviet-dupuis}
\end{thebibliography}
\end{document}